\documentclass[twocolumn,tighten]{aastex63}
\pdfoutput=1 
\usepackage{amsmath,amstext}
\usepackage{apjfonts} 
\usepackage{afterpage}
\usepackage{xcolor}
\usepackage{threeparttable}

\usepackage[encapsulated]{CJK}
\usepackage{ucs}
\usepackage[utf8x]{inputenc}
\newcommand{\cntext}[1]{\begin{CJK}{UTF8}{gbsn}#1\end{CJK}}

\usepackage{graphicx}
\usepackage{epsfig}

\usepackage{booktabs}
\usepackage{multirow}
\usepackage{float}
\usepackage{mathrsfs}
\usepackage{lineno}
\usepackage[symbol]{footmisc}
\renewcommand{\thefootnote}{\fnsymbol{footnote}}

\setwatermarkfontsize{100px} 

\shorttitle{Variability of the Black-Hole Image in M87}
\shortauthors{Satapathy et al.}

\begin{document}


\title{The Variability of the Black-Hole Image in M87 at the Dynamical Time Scale}

\author{Kaushik Satapathy}
\affiliation{Astronomy Department, University of Arizona, 933 N. Cherry Ave, Tucson, AZ 85721, USA}
\affiliation{Department of Physics, University of Arizona, 1118 E 4th St, Tucson, AZ 85721, USA}

\author{Dimitrios Psaltis}
\affiliation{Astronomy Department, University of Arizona, 933 N. Cherry Ave, Tucson, AZ 85721, USA}

\author{Feryal \"Ozel}
\affiliation{Astronomy Department, University of Arizona, 933 N. Cherry Ave, Tucson, AZ 85721, USA}

\author{Lia Medeiros}
\altaffiliation{NSF Astronomy and Astrophysics Postdoctoral Fellow}
\affiliation{School of Natural Sciences, Institute for Advanced Study, 1 Einstein Drive, Princeton, NJ 08540, USA}

\author{Sean T.\ Dougall}
\affiliation{Astronomy Department, University of Arizona, 933 N. Cherry Ave, Tucson, AZ 85721, USA}

\author{Chi-Kwan Chan}
\affiliation{Astronomy Department, University of Arizona, 933 N. Cherry Ave, Tucson, AZ 85721, USA}

\author{Maciek Wielgus}
\affiliation{Black Hole Initiative at Harvard University, 20 Garden Street, Cambridge, MA 02138, USA}
\affiliation{Center for Astrophysics | Harvard \& Smithsonian, 60 Garden Street, Cambridge, MA 02138, USA}

\author{Ben S.\ Prather}
\affiliation{Illinois Center for Advanced Studies of the Universe, Department of Physics, University of Illinois, 1110 West Green Street, Urbana, IL 61801, USA}

\author{George N.\ Wong}
\affiliation{Illinois Center for Advanced Studies of the Universe, Department of Physics, University of Illinois, 1110 West Green Street, Urbana, IL 61801, USA}

\author{Charles F.\ Gammie}
\affiliation{Illinois Center for Advanced Studies of the Universe, Department of Physics, University of Illinois, 1110 West Green Street, Urbana, IL 61801, USA}
\affiliation{Department of Astronomy, University of Illinois, 1002 West Green Street, Urbana, IL 61801, USA}

% ...

\begin{abstract}
The black-hole images obtained with the Event Horizon Telescope (EHT) are expected to be variable at the dynamical timescale near their horizons. For the black hole at the center of the M87 galaxy, this timescale (5--61 days) is comparable to the 6-day extent of the 2017 EHT observations. Closure phases along baseline triangles are robust interferometric observables that are sensitive to the expected structural changes of the images but are free of station-based atmospheric and instrumental errors. We explored the day-to-day variability in closure phase measurements on all six linearly independent non-trivial baseline triangles that can be formed from the 2017 observations. We showed that three triangles exhibit very low day-to-day variability, with a dispersion of $\sim3-5^\circ$. The only triangles that exhibit substantially higher variability ($\sim90-180^\circ$) are the ones with baselines that cross visibility amplitude minima on the $u-v$ plane, as expected from theoretical modeling. We used two sets of General Relativistic magnetohydrodynamic simulations to explore the dependence of the predicted variability on various black-hole and accretion-flow parameters. We found that changing the magnetic field configuration, electron temperature model, or black-hole spin has a marginal effect on the model consistency with the observed level of variability. On the other hand, the most discriminating image characteristic of models is the fractional width of the bright ring of emission. Models that best reproduce the observed small level of variability are characterized by thin ring-like images with structures dominated by gravitational lensing effects and thus least affected by turbulence in the accreting plasmas.
\end{abstract}


\section{Introduction}

The Event Horizon Telescope (EHT) has produced horizon scale images of the black hole in the center of the M87 galaxy using Very Long Baseline Interferometry (VLBI) at a wavelength of $1.3\ \rm{mm}$~\citep{2019ApJ...875L...1E,2019ApJ...875L...2E,2019ApJ...875L...3E,2019ApJ...875L...4E,2019ApJ...875L...5E,2019ApJ...875L...6E}. Together with upcoming studies of the black hole in the center of the Milky Way, Sgr A*, these images have opened up numerous avenues to studying physics at event-horizon scales related to both accretion processes and gravity. The structure of the accretion disks and jets observed around the black holes will improve our understanding of the behaviour of plasma in these environments~\citep{2019ApJ...875L...5E,2019ApJ...875L...6E}. The measurement of the sizes and shapes of the black-hole shadows gives insights into deviations of the black hole spacetime from the Kerr metric~\citep{2019GReGr..51..137P,2020PhRvL.125n1104P}. 

Accretion flows, which make up the horizon-scale environments of black holes, are expected to be highly variable by nature. The radiation observed from these regions at $1.3\ \rm{mm}$ is dominated by synchrotron emission from electrons which are heated and accelerated by magnetohydrodynamic (MHD) turbulence~\citep{2019ApJ...875L...5E}. The accretion flow is also expected to evolve at the dynamical time scales near the innermost stable circular orbit (ISCO), which range from 4 to 56 minutes for Sgr A* (for $M_{\rm{BH}} = 4.148 \times 10^6 M_\odot$; see \citealt{2019A&A...625L..10G}) and 5 to 61 days for M87 (depending on the black hole spin;~\citealt{Bardeen1972}). The timescales suggest that there can be substantial variability in the source structure over a single observing night for Sgr A*, which makes it difficult to obtain a static image. However, we expect substantial source variability in M87 only for observations that span about a week or longer. Evidence for long-term variability in the image structure of M87 has been reported recently, based on VLBI observations spanning nearly a decade~\citep{Wielgus2020}.

The EHT observed M87 in April 2017 for four nights with a maximum separation of six calendar days between observations. For a black hole of the inferred mass ($M_{\rm{BH}} = 6.5 \times 10^9 M_\odot$; \citealt{Gebhardt2011, 2019ApJ...875L...6E}), six days correspond to $16\ GM/c^3$. This is approximately equal to $\sim 3.5$ dynamical timescales at the ISCO for a maximally spinning black hole. Over this observation span, only a limited amount of variability in source structure was inferred between days, reflected as a small change in the thickness or azimuthal brightness distribution of the observed bright emission ring  (\citealt{2019ApJ...875L...4E}). 

Our goal in this paper is to use General Relativistic Magneto-Hydrodynamic (GRMHD) simulations of accretion flows, followed by General Relativistic ray tracing and radiative transfer, with parameters that are relevant to the M87 black hole, in order to explore the short-term variability in the source structure as it is imprinted in the interferometric observables. The EHT, being an interferometer, measures complex visibilities, which denote the two dimensional Fourier transform of the source image on the sky. The measurements are typically decomposed into visibility amplitudes and phases (\citealt{2017isra.book.....T}). Image reconstruction techniques enable us to use these variables and create the map of the source brightness on the plane of the sky (see \citealt{2019ApJ...875L...4E}).

The visibility measurements in VLBI are affected by time-dependent atmospheric and instrumental errors of the telescopes, particularly challenging for the short radio wavelengths (\citealt{2017isra.book.....T}). The custom-built EHT data reduction pipeline addresses the specific requirements of the millimeter VLBI (\citealt{2019ApJ...875L...3E}). The pipeline uses Atacama Large Millimeter/submillimeter Array (ALMA) as an anchor station, leveraging its extreme sensitivity for the calibration of the entire EHT array (\citealt{2019ApJ...882...23B},\citealt{2019A&A...626A..75J}). As a result, the measured visibilities indicate sufficient phase-stability to enable long coherent averaging and building up the high signal to noise ratio. Remaining station-based errors can be modelled as complex gains multiplying the visibilities. In order to eliminate the effects of these errors in the data, one can construct closure quantities with amplitudes and phases (\citealt{1958MNRAS.118..276J}, \citealt{1989AJ.....98.1112K}, \citealt{2017isra.book.....T}, \citealt{2020ApJ...894...31B}). For a set of three baselines forming a closure triangle, the closure phase is defined as the sum of the measured visibility phases in each of these baselines. This cancels out the station-based phase measurement errors for each station. As a result, closure phases are optimal and robust probes of the intrinsic structure of the source.

Closure phases have indeed proven to be a powerful tool to study and quantify variability in synthetic data from GRMHD simulations. \citet{2018ApJ...856..163M} studied the dependence of variabililty in closure phases on the orientations and baseline lengths of the closure triangles (see also~\citealt{Roelofs2017}). Triangles that involve large baselines (that probe small length scales in the source image) were shown to have a high degree of variability in closure phases. On the other hand, triangles that involve small baselines (that probe the overall size of the ring and source structure) exhibit less variability. The exception to this rule are triangles involving a baseline close to a deep visibility minimum. The visibility phases in regions of visibility minima are extremely sensitive to minor changes in source structure. This localization of variability in the Fourier space is vital in understanding variability in GRMHD simulations and in observations. 

In this paper, we first present a data-driven model to quantify the variability of observed closure phases in the various linearly independent triangles of the M87 observations across the six days of observations in 2017. We apply this algorithm to the 2017 EHT data on M87 and identify three closure triangles that show a remarkably small degree of variability. We then explore the degree of variability produced in a large set of GRMHD simulations, with different magnetic field configurations and prescriptions of the plasma physics, and understand the effects of various model parameters on the variability in the models. Finally, we compare the predictions of the GRMHD simulations to the observations and discuss how the latter constrain the physical properties of the accretion flow in M87.

\begin{figure*}[t]
	\centerline{\includegraphics[height = 0.419\textwidth, width =  0.419\textwidth]{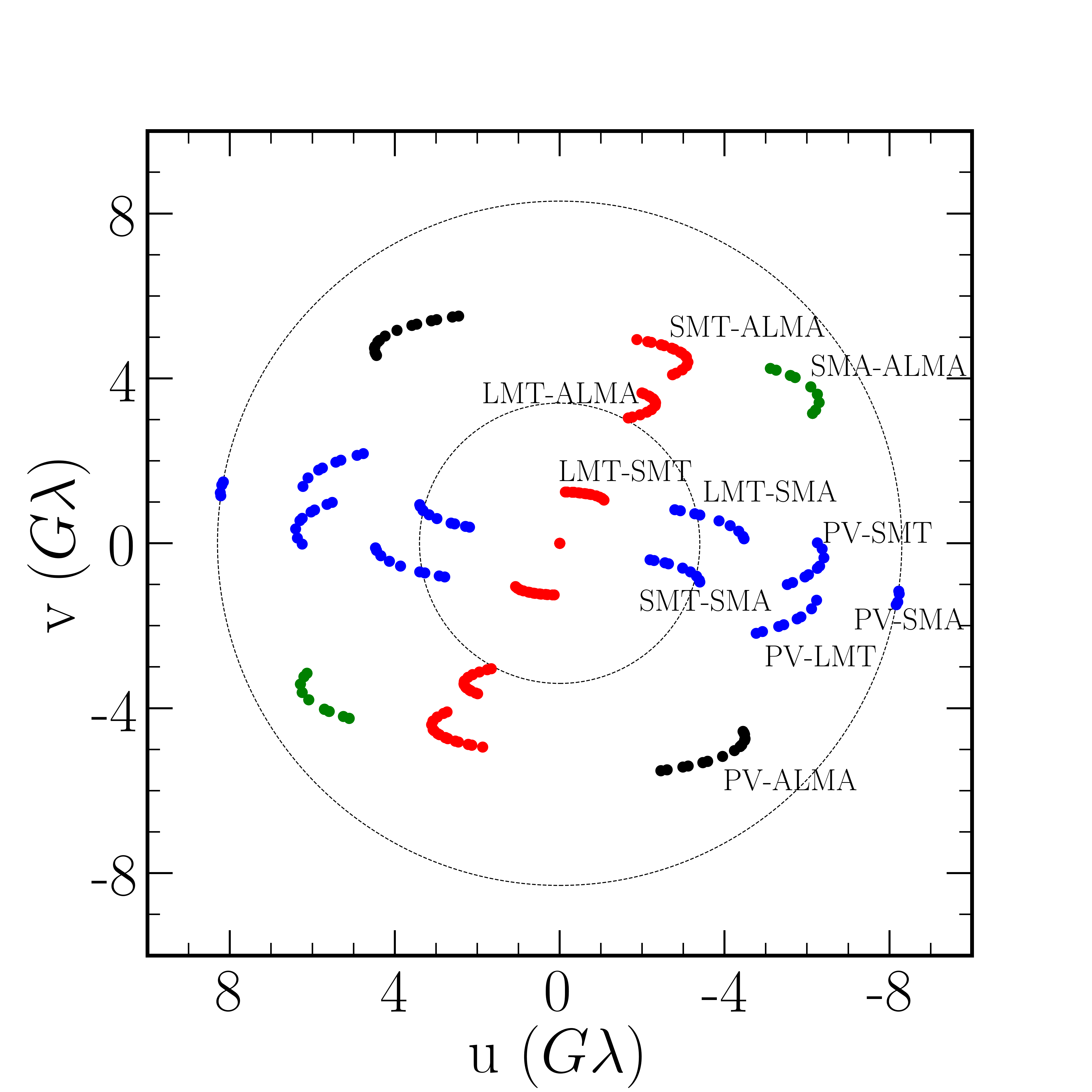}
				\includegraphics[height = 0.419\textwidth, width =  0.580\textwidth]{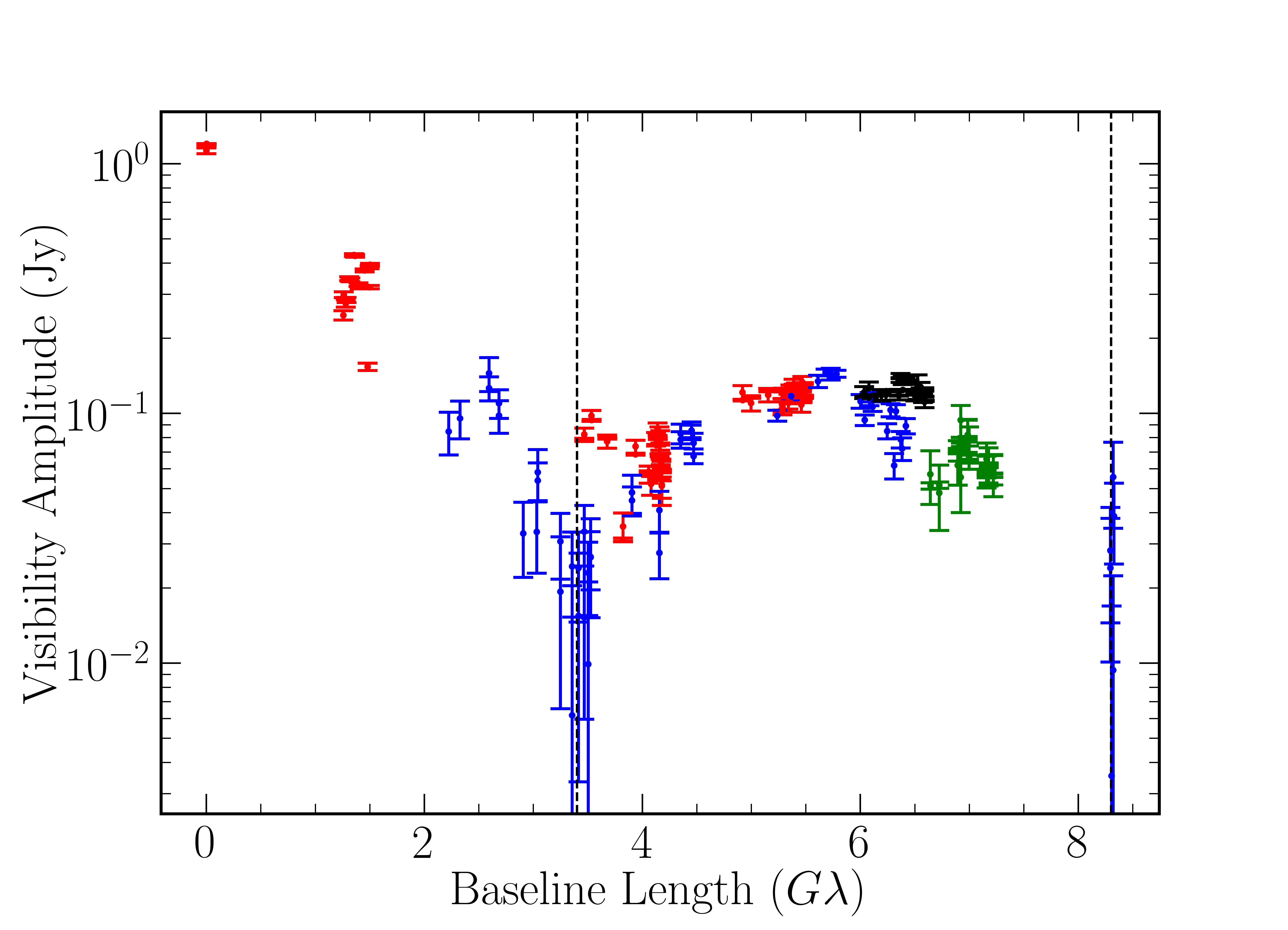}} 
\caption{\textit{(Left)} : The $u-v$ coverage traced by the EHT for the M87 observations on 2017, April 11, with the dashed circles indicating locations of the two observed visibility minima. The colors indicate different orientations of the $u-v$ coverage. \textit{(Right)}: The visibility amplitude as a function of baseline length on the same day.}
\label{fig:uv-coverage}
\end{figure*}

\section{Closure Phase Observations of M87}
\label{sec:section 2}

The EHT measures complex visibilities, given by 
\begin{equation}
\label{two-dim-fft}
V(u,v) = \int\int dx\ dy\ I(x,y) \ e^{ -2\pi i (ux + vy) },
\end{equation}
\noindent where $I(x,y)$ represents the total intensity of radiation at a given spatial location $(x,y)$ on the image plane and $(u,v)$ denote the Fourier frequencies corresponding to the $(x,y)$ coordinates. However, due to atmospheric and instrumental effects, the measured visibilities do not represent the actual visibilities of the source. The measured visibilities are related to the source visibilities through complex "gains". The measured visibility between the $i$-th and the $j$-th telescopes, $\mathscr{V}_{ij}$, can be written as 

\begin{equation}
\label{eq:vis_gains}
\mathscr{V}_{ij} = g_i g^*_j V_{ij},
\end{equation} 

\noindent where $g_i$ and $g_j$ are the complex gains associated with the two telescopes and the star superscript indicates complex conjugates. We define the closure phase as the argument of the complex bispectrum along a closed triangle of three telescopes (\citealt{1958MNRAS.118..276J}). The closure phase $\phi^{\rm{CP}}_{ijk}$ for a closure triangle formed by the $i$-th, the $j$-th, and the $k$-th telescopes is hence given by

\begin{align}
\label{eq:cp_equation}
\begin{split}
\phi^{\rm{CP}}_{ijk} &= \rm{{Arg}}[\mathscr{V}_{ij}\mathscr{V}_{jk}\mathscr{V}_{ki}]\\ &= \rm{Arg}[g_i g^*_j V_{ij} g_j g^*_k V_{jk} g_k g^*_i V_{ki}]\\
&= \rm{Arg}[|g_i|^2 |g_j|^2 |g_k|^2 V_{ij} V_{jk} V_{ki}] \\
&= \rm{Arg}[V_{ij} V_{jk} V_{ki}]\\
&= \phi_{ij} + \phi_{jk} + \phi_{ki}, 
\end{split}
\end{align}
where $\phi_{ij}$ denotes the visibility phase between the $i$-th and the $j$-th telescopes. Equation \ref{eq:cp_equation} shows that the measured closure phase in a triangle is equal to the actual closure phase represented by the source. 

The EHT observed M87 on four nights in 2017 - April 5, April 6, April 10, and April 11. The observations involved seven stations spread across five geographical locations (\citealt{2019ApJ...875L...2E}): the Atacama Large Millimeter Array (ALMA) and the Atacama Pathfinder Experiment (APEX) in Chile; the James Clerk Maxwell Telescope (JCMT) and the Submillimeter Array (SMA) in Hawai'i; the Large Millimeter Telescope (LMT) in Mexico; the Submillimeter Telescope Observatory (SMT) in Arizona; and the IRAM 30m telescope on Pico Veleta (PV) in Spain. On each night of observation, the rotation of the earth implies that each baseline traces out an elliptical path in the $u-v$ space with the progression of time. 

Figure \ref{fig:uv-coverage} shows the $u-v$ coverage for M87 during one of the observing days (left panel) as well as the visibility amplitudes measured as a function of baseline length on the same day (right panel). Two deep minima in visibility amplitude can be seen at baseline lengths of $\sim 3.4 G\lambda$ and $\sim 8.3 G\lambda$ encountered in three baselines (LMT-SMA, SMT-SMA, PV-SMA) along the east-west orientation (indicated in blue). In contrast, the minimum encountered by the ALMA-LMT baseline is only marginally deep (right panel of Figure \ref{fig:uv-coverage}). 

We construct closure phase quantities using all the baselines shown in Figure \ref{fig:uv-coverage}. However, closure triangles with two telescopes at the same geographical location yield trivial closure phases, with deviations from zero arising due to changes in the large scale source structure alongside systematic and thermal errors (\citealt{2019ApJ...875L...3E}). While these triangles are useful to quantify systematic errors in closure phases, they do not capture any information about the source structure at horizon scales. Hence, we only choose triangles with telescopes at different geographical locations for our study. Since Chile and Hawai'i have two stations each, we pick the station with the higher signal-to-noise ratio.  We, therefore, consider five stations - ALMA, SMA, LMT, SMT, and PV, out of which six linearly independent closure phases can be constructed. We list the six triangles in Table \ref{tab:closure triangles}. We choose the triangles such that all of them involve ALMA, which has the highest signal-to-noise ratio, and hence smaller uncertainties in closure phase measurements.

\begin{table*}[t]

\caption{Closure Triangles used with Baseline Lengths and Inferred Variability in Closure Phases.}
\begin{center}
\begin{threeparttable}
\begin{tabular}{ c c c c c}
\toprule
\multirow{2}{*}{Triangle\tnote{a}} & \multicolumn{3}{c}{Baseline Lengths (G$\lambda$)\tnote{b}} & \multirow{2}{*}{Inferred Variability\tnote{c}} \\
\cmidrule{2-4}
 & 1 & 2 & 3 \\
 \midrule
\textbf{ALMA - SMA - LMT} & 6.7 -- 7.2 & \textbf{3.1 -- 4.5} & \textbf{3.4 -- 4.1} & $\sim30^\circ-60^\circ$ \\
ALMA - LMT - SMT & 3.4 -- 4.2 & 1.3 -- 1.5 & 4.8 -- 5.5 & $3.4^\circ$ \\
ALMA - PV - LMT & 6.0 -- 6.6 & 5.1 -- 6.4 & 3.4 -- 4.2 & $4.6^\circ$ \\
ALMA - \textbf{SMA - SMT} & 6.7 -- 7.2 & \textbf{2.4 -- 3.5} & 4.8 -- 5.5 & $\sim180^\circ$\\
ALMA - \textbf{PV - SMA} & 6.0 -- 6.2 & \textbf{8.2 -- 8.3} & 6.8 -- 7.0 & -\\
ALMA - PV - SMT & 6.0 -- 6.6 & 5.5 -- 6.4 & 5.3 -- 5.5 & $5.4^\circ$ \\
\bottomrule
\end{tabular}
\begin{tablenotes}
\item \textbf{Notes}
\item[a] The bold baselines indicate the triangles which cross one of the deep visibility minima in the E-W orientation at 3.4 G$\lambda$ and 8.3 G$\lambda$.
\item[b] Baselines labeled 1, 2, and 3 in a triangle labelled as A-B-C correspond to the baselines A-B, B-C. and C-A respectively.
\item[c] An estimated range of variation in the high-variability triangles, and the maximum likelihood value of the inferred Gaussian variability (Equation \ref{eq:likelihood defn}) in the low-variability triangles 
\end{tablenotes}
\end{threeparttable}
\end{center}
\label{tab:closure triangles}
\end{table*}

\begin{figure}[t]
\includegraphics[width=0.5\textwidth]{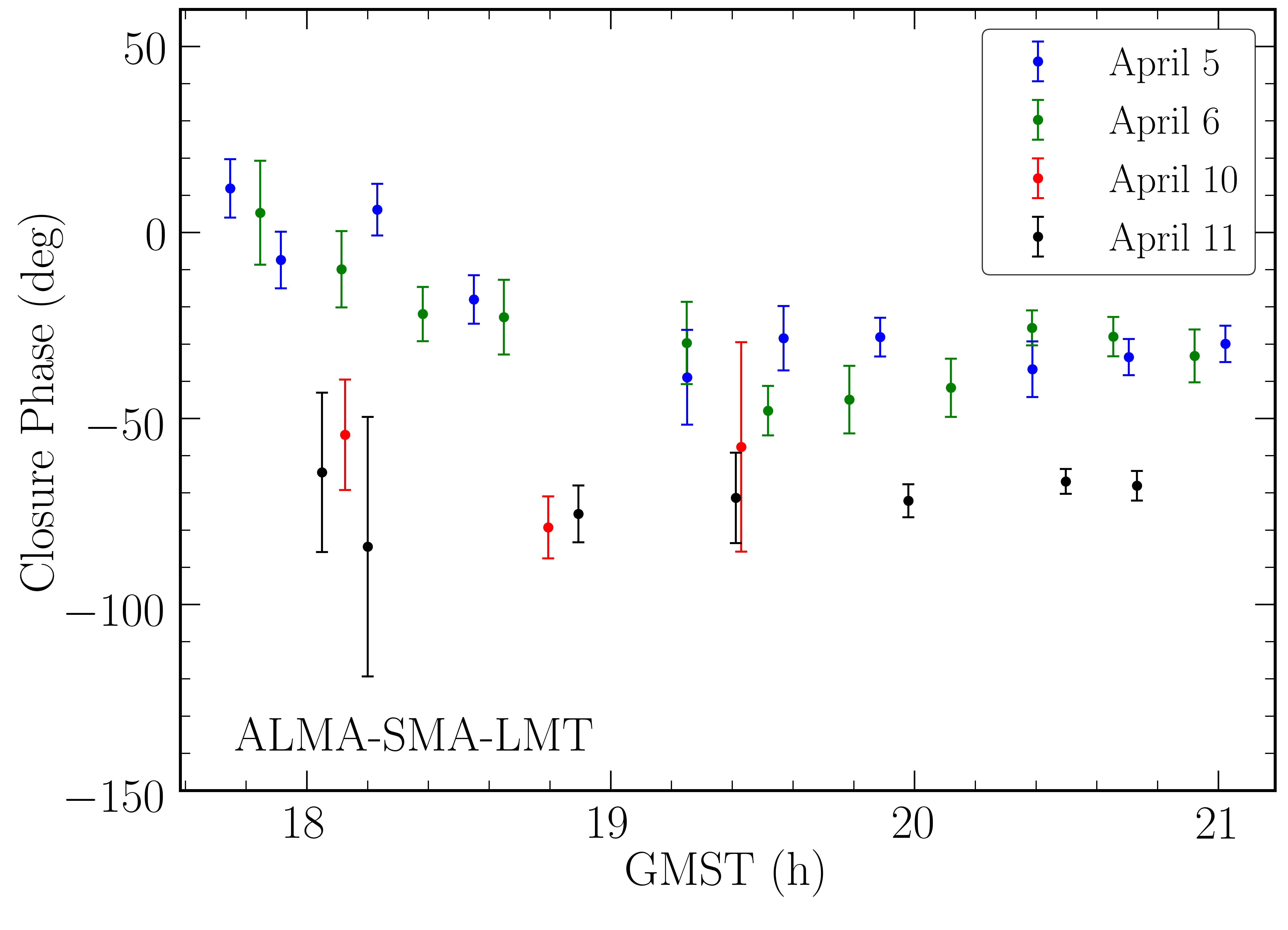}
\includegraphics[width=0.5\textwidth]{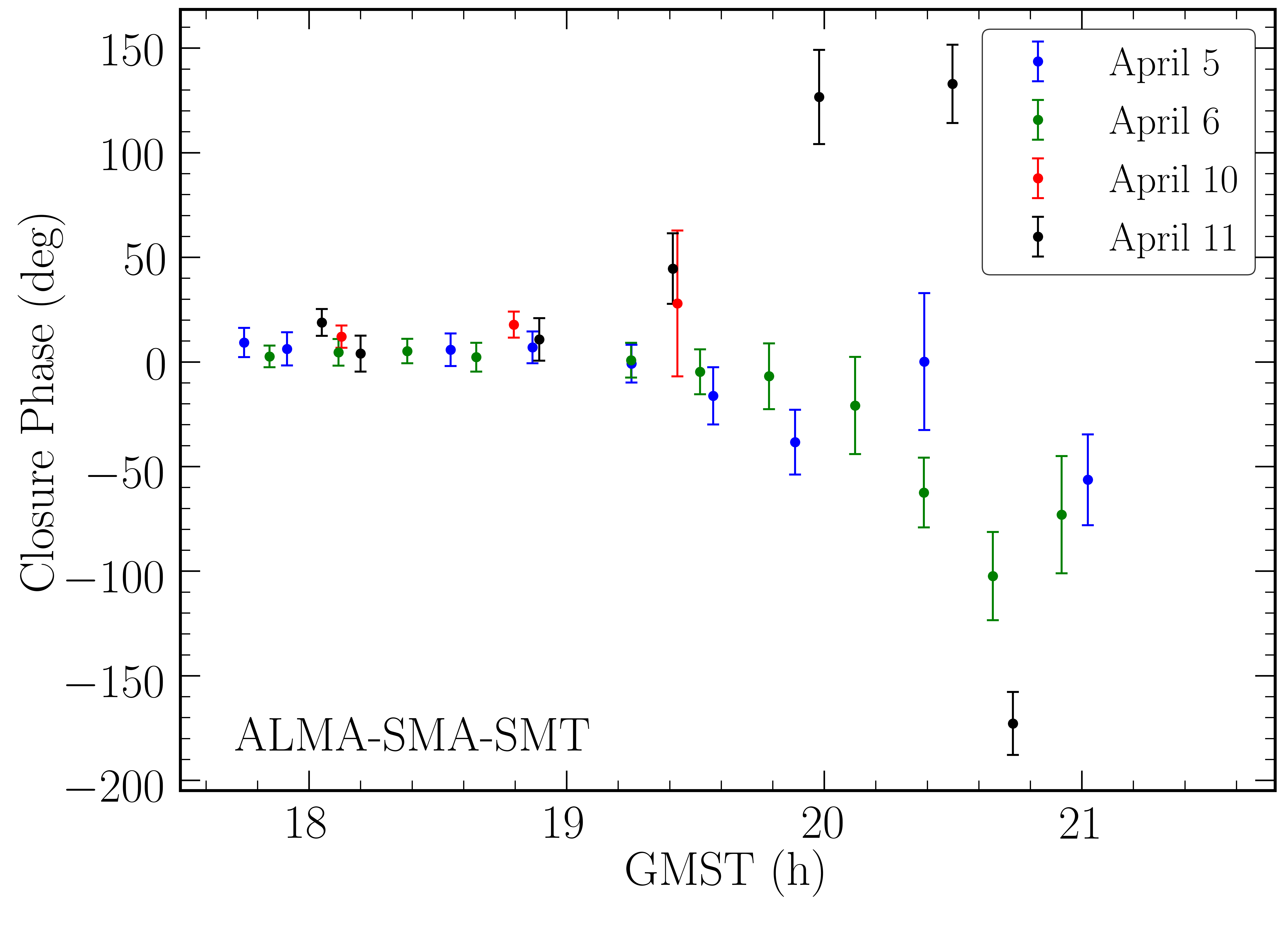}
\includegraphics[width=0.5\textwidth]{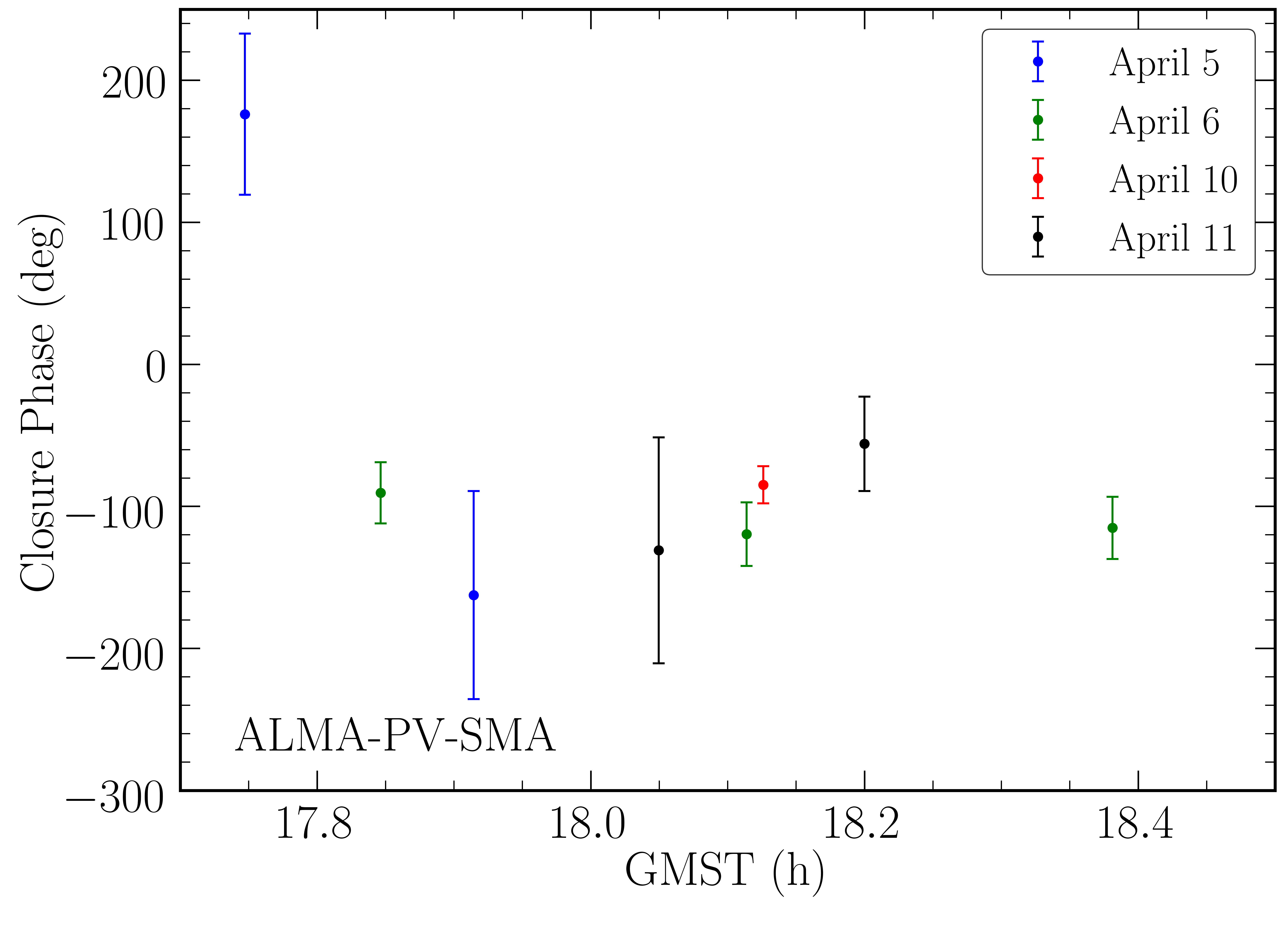}
\caption{The evolution of closure phase plotted with time for all four days of observation for closure triangles in which baselines are known to encounter regions of deep visibility minima. All the three triangles exhibit a high level of variability in closure phases across six days of observation.\\ }
\label{fig:variable-triangles}
\end{figure}

\begin{figure*}[t]
	\centerline{\includegraphics[width=0.5\textwidth]{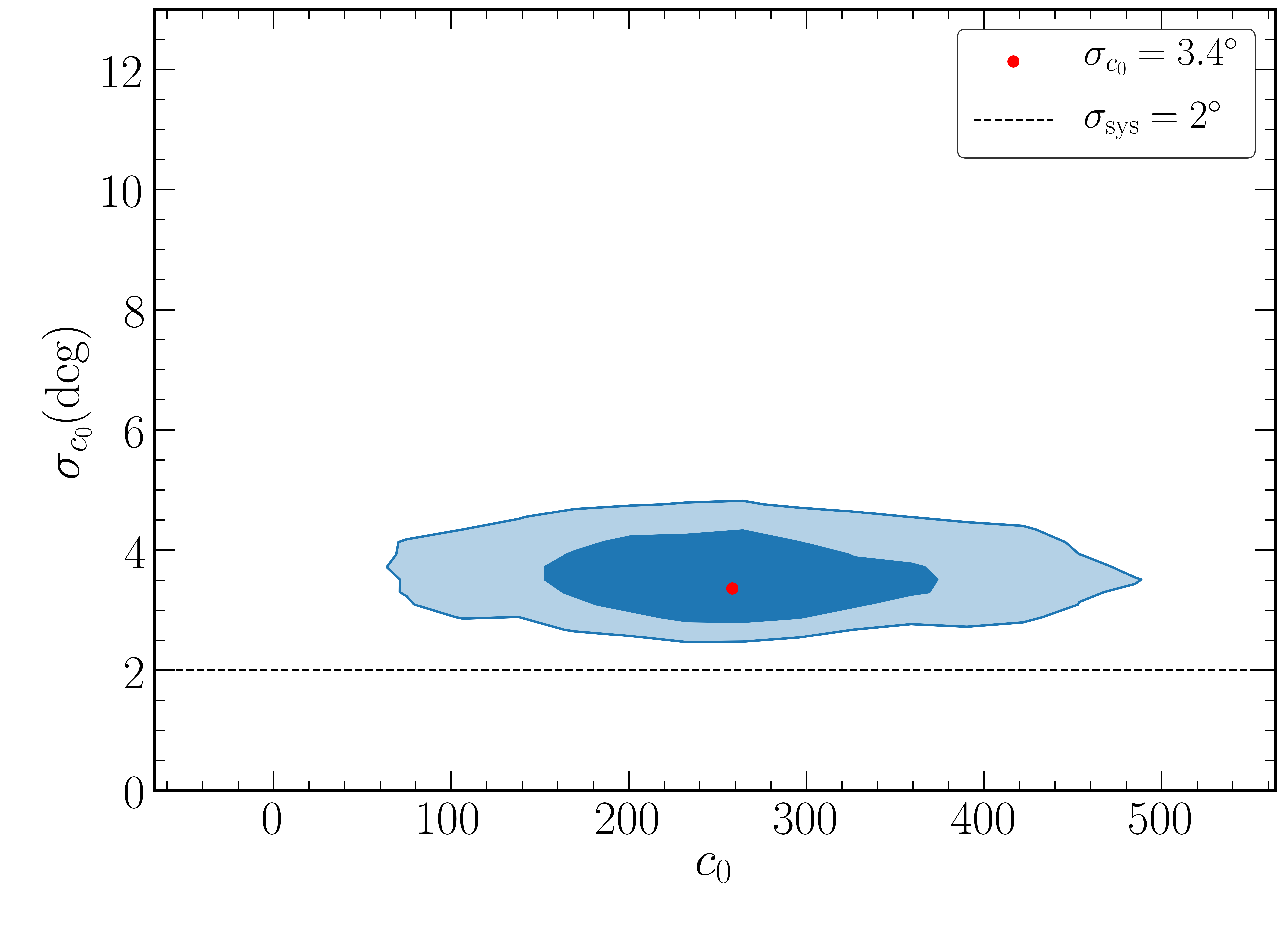}
				\includegraphics[width=0.5\textwidth]{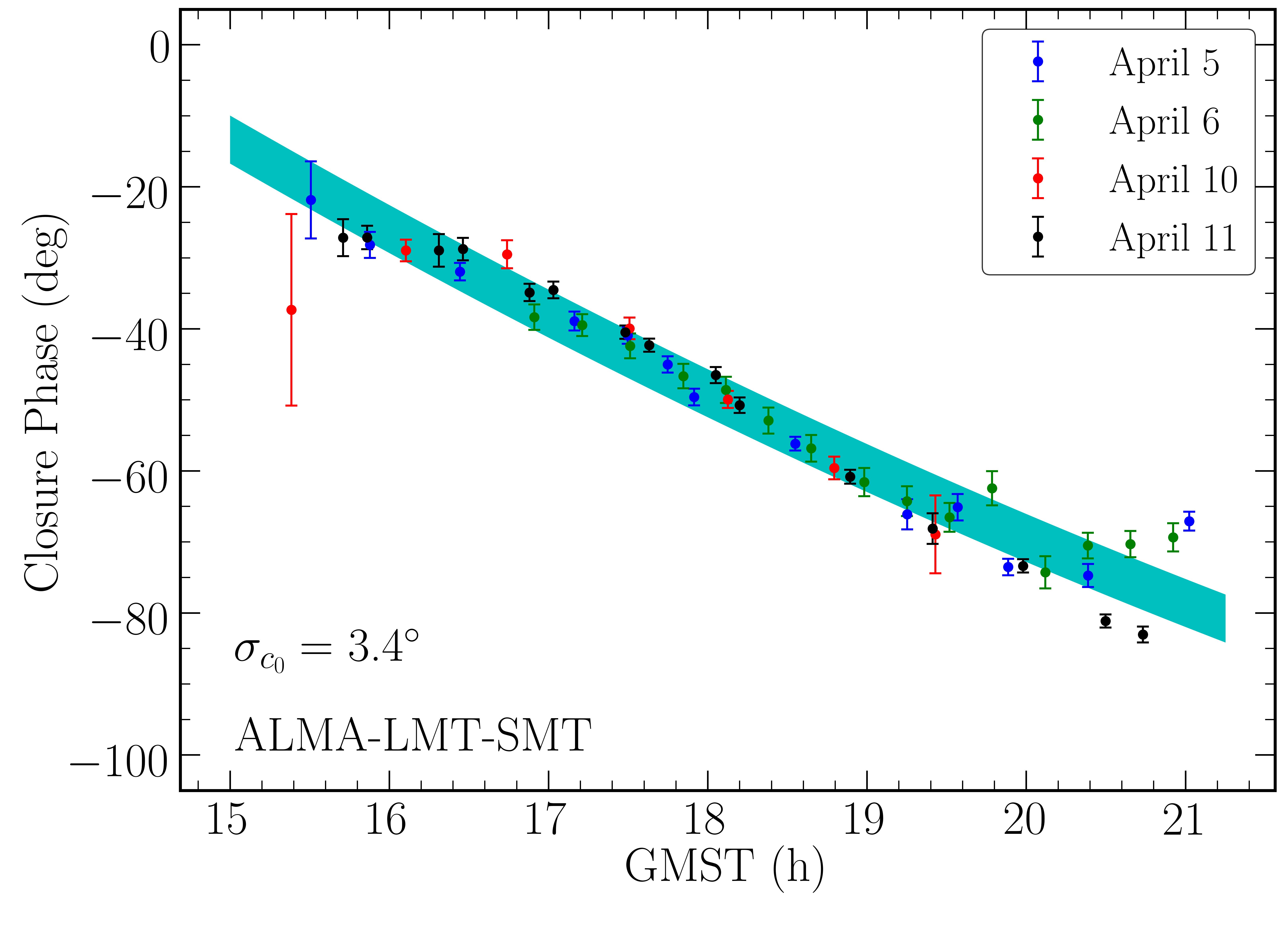}}
	\centerline{\includegraphics[width=0.5\textwidth]{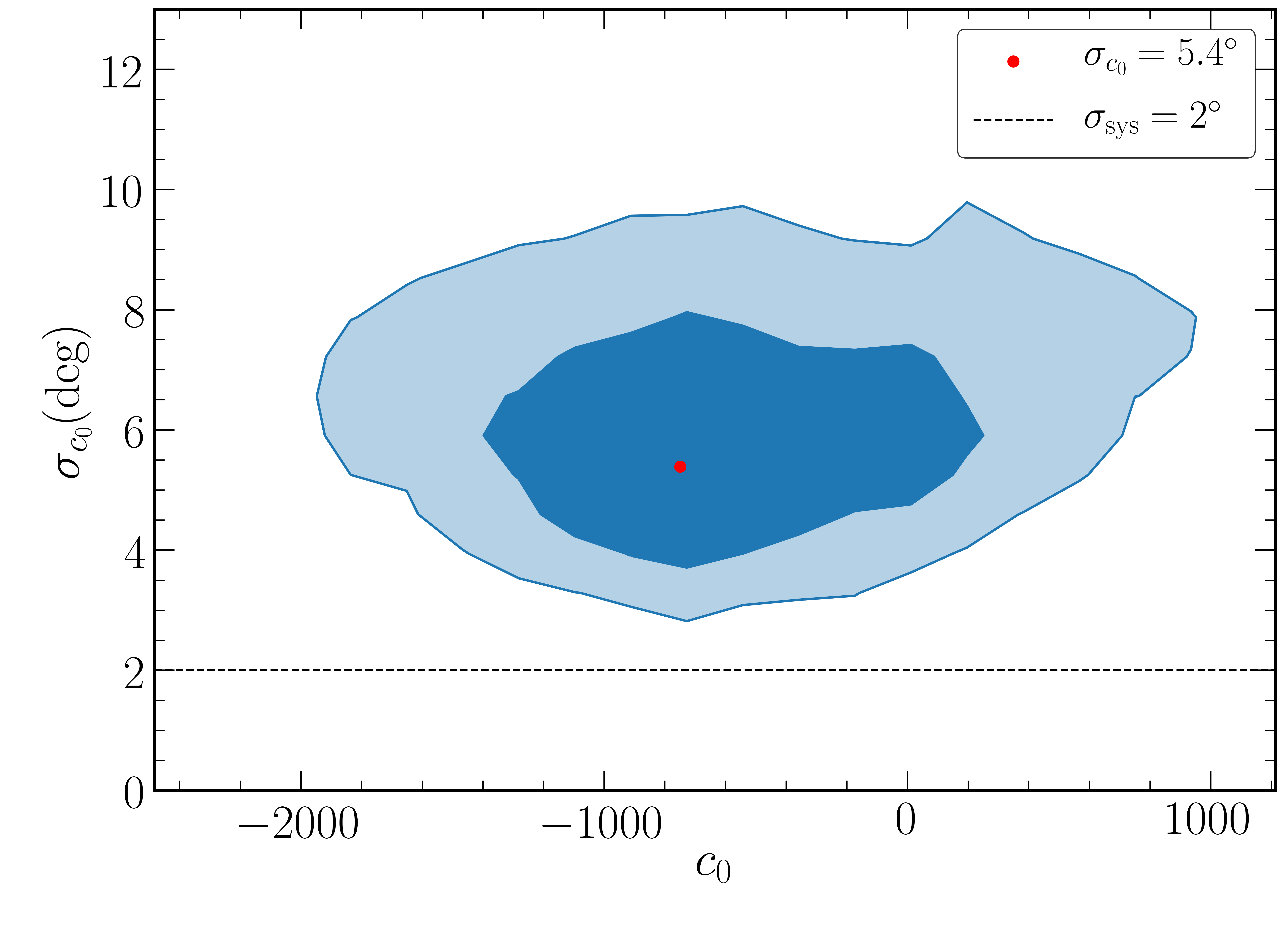}
				\includegraphics[width=0.5\textwidth]{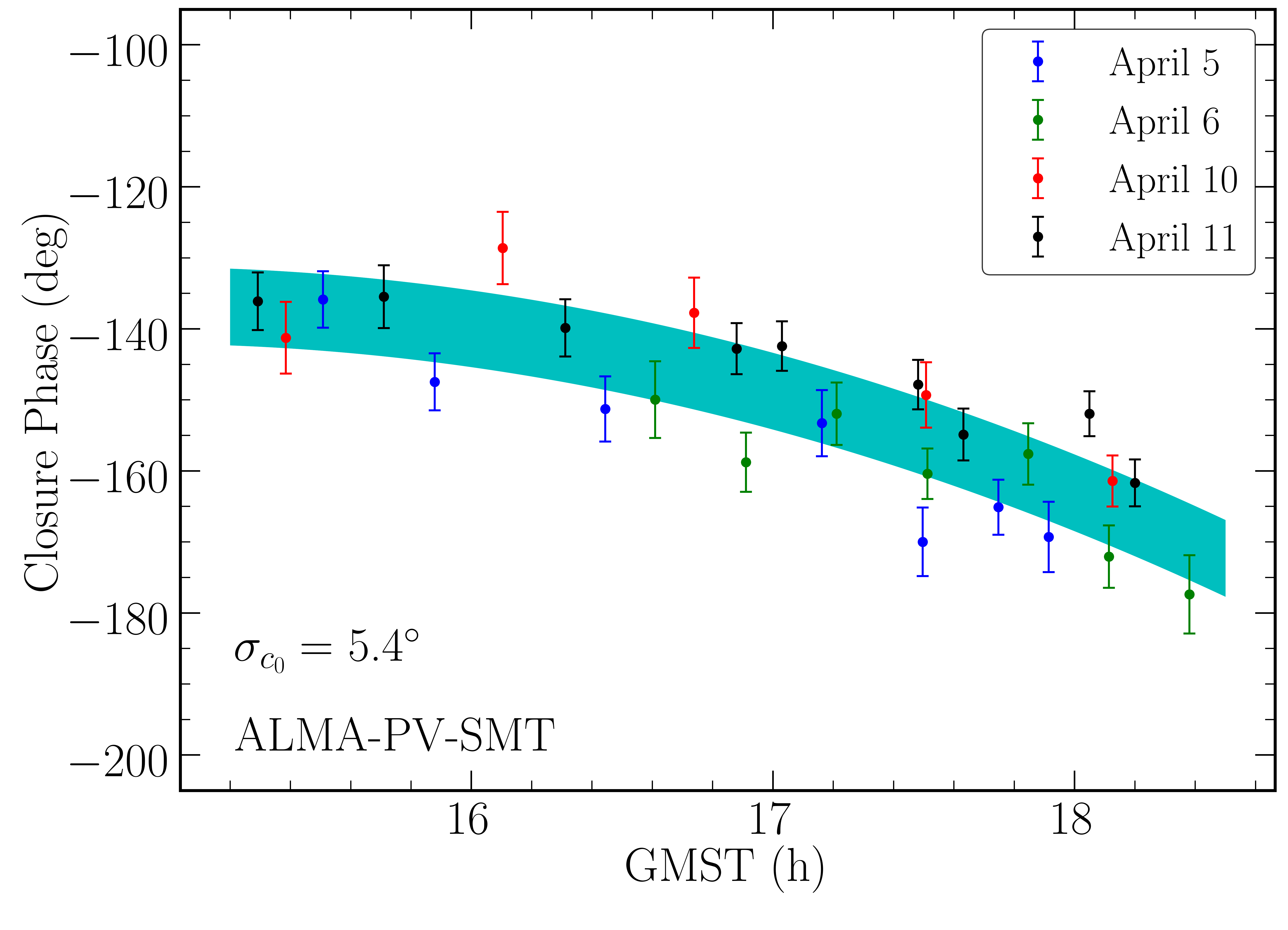}}  
	\centerline{\includegraphics[width=0.5\textwidth]{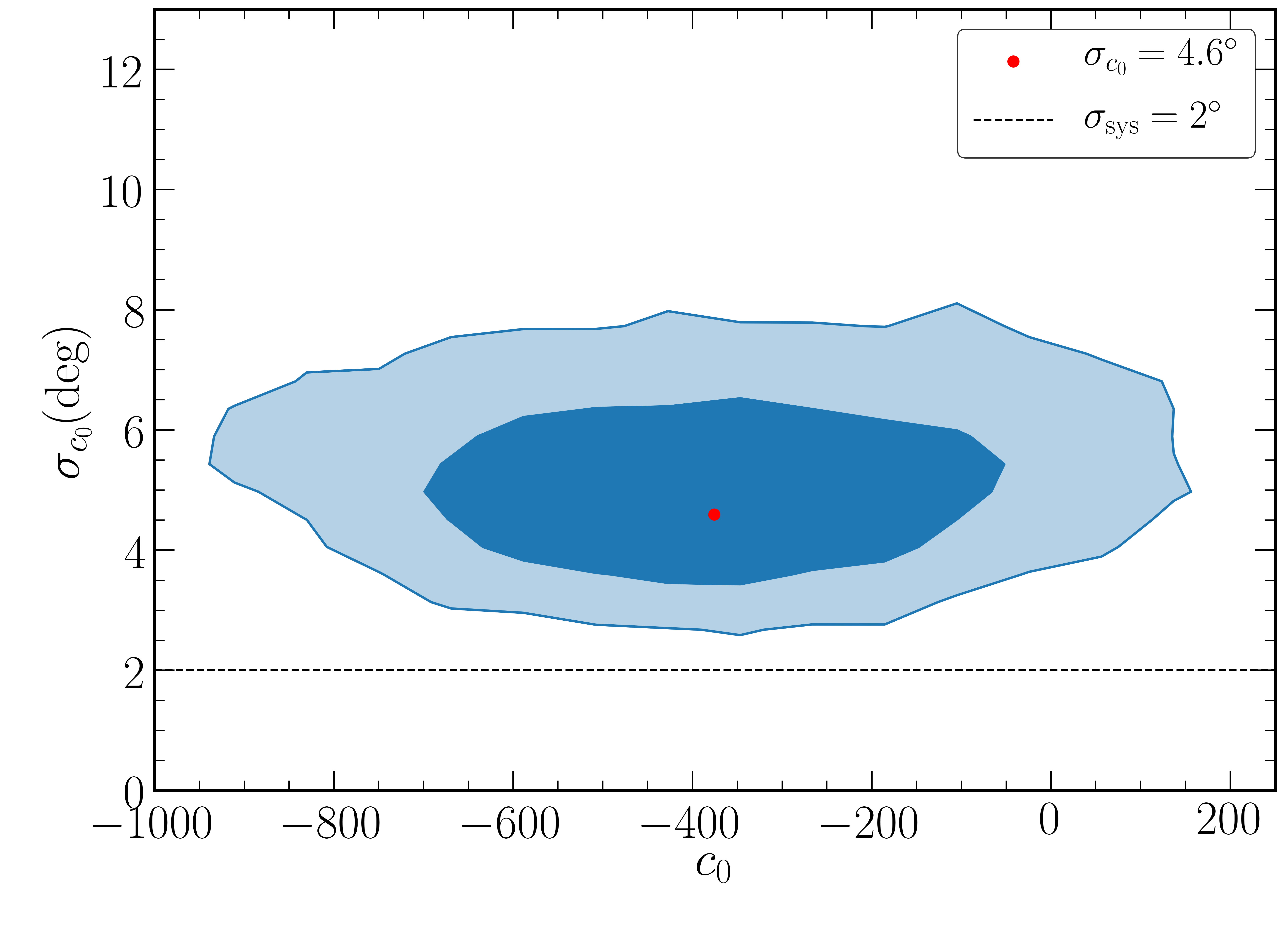}
				\includegraphics[width=0.5\textwidth]{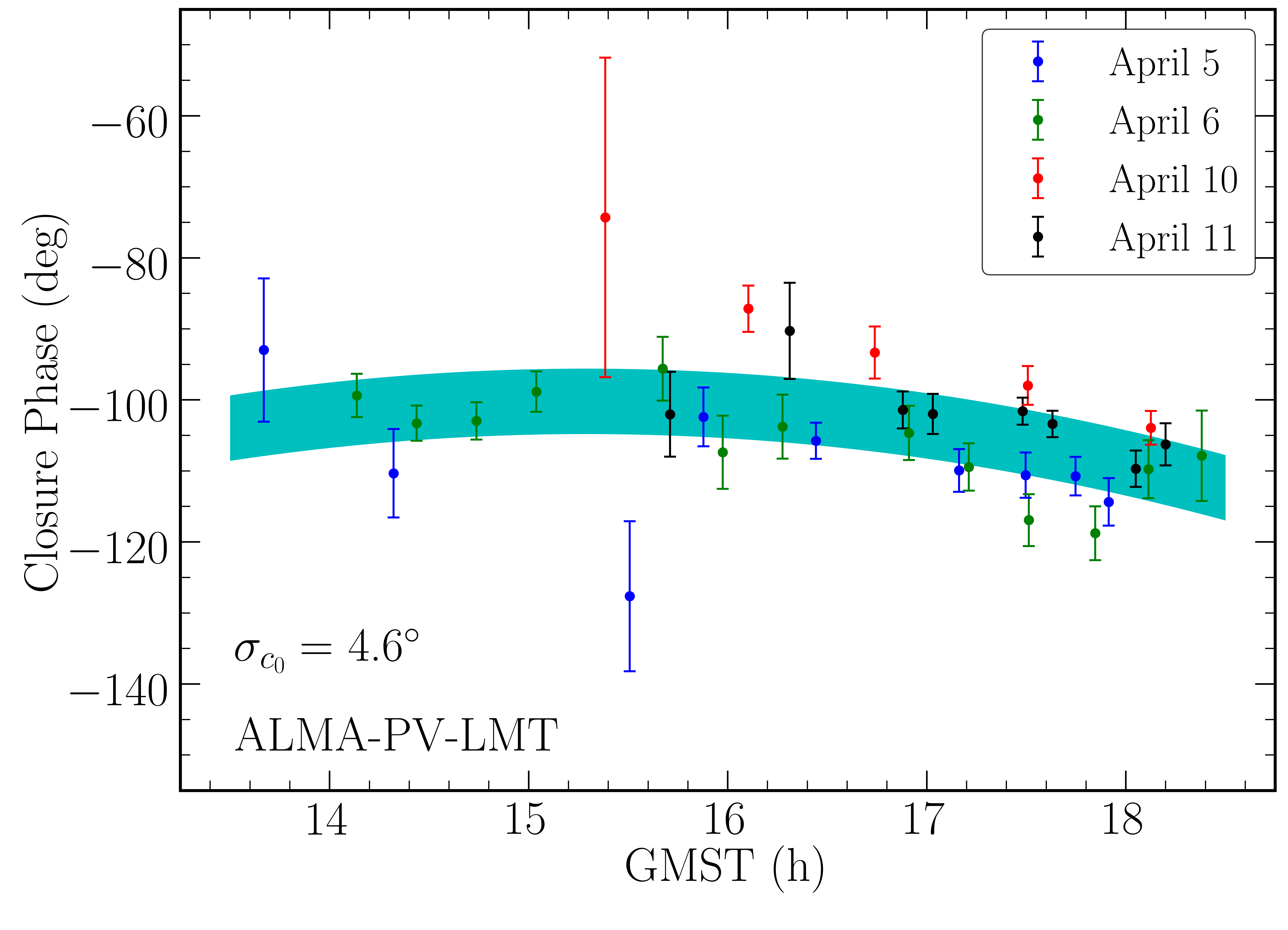}}			
\caption{The parameters and functional forms of the data-driven closure phase evolution model, applied to the closure triangles that exhibit low variability across the 6 days of 2017 EHT obsevations (see eq.~\ref{eq:quadratic model}). \textit{(Left)}: The posterior projected on the $\sigma_{c_0}-c_0$ parameter plane, where $\sigma_0$ is a measure of the variability across the 6 days. The red dots indicate the most likely values. The contours indicate the levels containing 68\% and 95\% of the posterior probability. The dashed horizontal lines indicate the systematic error of $2^\circ$ in closure phase observations.  \textit{(Right)}: Closure phase observations plotted as a function of time for the same triangles. The cyan band indicates the most likely value of the width of the model $\sigma_{c_0}$. }
\label{fig:lc-variability}
\end{figure*}

In Figure~\ref{fig:variable-triangles}, we show the evolution of closure phases with time in Greenwich Mean Sidereal Time (GMST) coordinates over all four days of observations in three triangles that exhibit substantial day-to-day variability in closure phases over the span of the 2017 campaign (see also \citealt{2019ApJ...875L...3E} Section 7.3.2):\\
\noindent (\textit{i}) The ALMA-SMA-LMT triangle shows a change of $~30^\circ-60^\circ$ between the first two days and the later two days of observation. \\
\noindent (\textit{ii}) The ALMA-SMA-SMT triangle shows a swing of $~180^\circ$ on each day of observation. In addition to this, the direction of this swing flips between the first two days and the later two days of observation.  \\
\noindent (\textit{iii}) In the ALMA-PV-SMA triangle, it is difficult to quantify the degree of variability because of the reduced complex visibility amplitudes between PV and SMA baselines, and the low signal-to-noise ratio in some data points. However, a persistent structure in the evolution of closure phases is not observed. 

\noindent We note that each of these closure triangles involve one of the baselines (i.e., LMT-SMA, SMT-SMA, PV-SMA) that cross a deep visibility minimum in the E-W orientation. In addition, we note that in the ALMA-SMA-SMT triangle, the closure phases show little day-to-day variability until $\sim 19$ GMST on each day. We attribute this behavior with the fact that the SMT-SMA baseline encounters the deep visibility minimum at $\sim 19$ GMST. We plan to analyze the dependence of variability in the observations on the depth of the visibility minima in a future study. 

The closure phases in the remaining triangles (i.e., ALMA-LMT-SMT, ALMA-PV-LMT, and ALMA-PV-SMT) show a persistent and continuous evolution with time during each day of observation (see Figure \ref{fig:lc-variability}). This behaviour represents the evolution of closure phases as the various baselines trace their paths on the $u-v$ plane following the rotation of the Earth. However, the same evolution with time is repeated in all days of observations; there is little scatter around the general trend set by the rotation of the baselines. The presence of substantial scatter from day to day would have been a signature of structure variability in the image.

In order to compare the observations to theoretical models, we quantify the degree of closure phase day-to-day variability observed in this last set of triangles. Upon examination of the closure phases in the maximal set of closure triangles, we find that outside the set of linearly independent triangles that we choose, only one triangle (LMT-SMT-PV) shows a low degree of day-to-day variability. We conclude that it is optimal and sufficient to quantify the day-to-day variability in the low-variability triangles that we choose in Table \ref{tab:closure triangles}, since the LMT-SMT-PV triangle can be obtained by a linear combination of these three triangles.  

We also note that the comparison between observations and models could, in principle, be carried out for both high variability and low variability triangles. However, as we will discuss in $\S$ \ref{sec:sec4}, images from theoretical models uniformly show a high degree of phase variability in baselines that cross a visibility minimum and, as a result, a comparison with the high-variability triangles in the data turns out not to have much discriminating power between various models. The degree of variability exhibited on the low-variability triangles identified in the data, on the other hand, is a more significant challenge to the GRMHD models and proves to be a useful tool to distinguish between them.

The quantification of day-to-day variability in the set of three low-variable triangles is not straightforward, because closure phases evolve with time on each observing day but the individual scans on each day are not aligned at the exact same times. As a result, a difference in closure phase measured in two scans separated by almost (but not exactly) one sidereal day incorporates both the change due to the slightly different locations on the $u-v$ plane probed by the two scans and the change due the structural changes in the image.

In order to disentangle the two sources of variability, we employ a data-driven analytic model for the evolution of closure phase with time during a single day of observation. This is meant to capture the change in closure phases introduced by the changing location on the $u-v$ plane of the closure triangles. We then compare the overall change of the parameters of this model from day to day, in order to quantify the variability due to structural changes of the image.

For our data-driven model, $\phi_{\rm{model}}(t)$, we use a second order polynomial in the observation time $t$ (in GMST co-ordinates), given by
\begin{equation}
\phi_{\rm{model}}(t) = c_0 + c_1 t + c_2 t^2\;.
\label{eq:quadratic model}
\end{equation}

In order to account for potential structural changes from day to day, we allow for an intrinsic Gaussian spread in the zeroth order coefficient of this polynomial ($c_0$), with a standard deviation of $\sigma_{c_0}$. In this prescription, we do not assign any physical significance to the polynomial coefficients. Since the maximum separation between any two nights of observations is six days, $\sigma_{c_0}$ acts as a constraint on the overall change in closure phase at a given time in a given triangle across six days. 

Using this model, we define the likelihood of making a particular closure phase measurement using a Gaussian mixture model as 
\begin{equation}
\begin{split}
L_i(c_0,c_1,c_2,\sigma_{c_0}) \equiv \left\{ \frac{1}{\sqrt{2\pi(\sigma_i^2 + \sigma_{c_0}^2)}} \right\}\\ \times \exp \left[-\frac{1}{2}\frac{ \left\{\phi_{i} - \phi_{\rm{model}}(t_i)\right\}^2 }{\left(\sigma_i^2 + \sigma_{c_0}^2\right)} \right], 
\end{split}
\end{equation}
where $\phi_i$ is the closure phase in a triangle at an observation time $t_i$ and $\sigma_i$ denotes the uncertainty in $\phi_i$. Note that we have assumed here Gaussian statistics for the measurements of closure phases. This assumption is not expected to introduce any significant changes in our quantitative results since the triangles we will be applying this method to are characterized by high signal-to-noise measurements. Applying this mixture model to data with large errors would require using an appropriate distribution for the closure-phase errors and their correlations (see~\citealt{Christian_2020}).

The likelihood that $n$ closure phase observations in a given triangle across all four days of observation follow the model $\phi_{\rm{model}}(t)$ is
\begin{gather}
L(c_0,c_1,c_2,\sigma_{c_0})  = \prod\limits_{i=1}^{n} L_i(c_0,c_1,c_2,\sigma_{c_0})  \\ 
= \prod\limits_{i=1}^{n} \left\{ \frac{1}{\sqrt{2\pi (\sigma_i^2 + \sigma_{c_0}^2)}} \right\} \exp \left[ - \sum\limits_{i=1}^n \frac{1}{2}\frac{\left\{ \phi_{i} - \phi_{\rm{model}}(t_i) \right\}^2}{(\sigma_i^2 + \sigma_{c_0}^2)} \right]. 
\label{eq:likelihood defn}
\end{gather}
\noindent Assuming flat priors in the polynomial coefficients and in $\sigma_{c_0}$, we map the parameter space using Markov-chain Monte Carlo (MCMC) sampling. 

Figure~\ref{fig:lc-variability} shows the posteriors over two of the model parameters, $c_0$ and $\sigma_{c_0}$, for each of the three triangles that exhibit low day-to-day variability. The left panels show the $68$\% and $95$\% contours of the posteriors; the right panels show the data-driven model with the most likely value of the parameters, overplotted on the closure phase observations in these triangles. The most likely standard deviations of variability across the 6 days of observations are $\sim3$--$5^\circ$. This is comparable to the systematic error in closure phase observations of $2^{\circ}$ as reported in \citet{2019ApJ...875L...3E} and is remarkably small. 

\section{GRMHD Simulations and Library}
\label{sec:GRMHD}

A large suite of GRMHD simulations has been generated to model the accretion flow around $\rm{M87}$. The simulations have been performed using the algorithms \texttt{harm} (see \citealt{Gammie_2003}), \texttt{BHAC} (\citealt{2018JPhCS1031a2008O}), \texttt{KORAL} (\citealt{2014MNRAS.439..503S}), etc. The simulations have been initialized with two magnetic field configurations that led to different field structures in the flows: SANE (Standard and Normal Evolution, see \citealt{2003ApJ...592.1042I}) and MAD (Magnetically Arrested Disk, see \citealt{2012MNRAS.426.3241N} and \citealt{2013MNRAS.436.3856S}). A comprehensive comparison of the GRMHD algorithms is presented in \citet{2019ApJS..243...26P}. The set of observables that result from these simulations have been obtained by general relativistic ray tracing and radiative transfer algorithms, such as \texttt{GRay} (\citealt{2013ApJ...777...13C}) and \texttt{ipole} (\citealt{2018MNRAS.475...43M}).

In this paper, we analyze the GRMHD simulations that were run using \texttt{harm} (see \citealt{Gammie_2003} and \citealt{2012MNRAS.426.3241N}). The simulations include SANE and MAD configurations of the magnetic fields and different black hole spins with co-rotating or counter-rotating accretion disks. The flow properties, as calculated in the GRMHD simulations, scale with the mass of the black hole and, therefore, the latter does not enter the calculation. The dependence of the electron temperature on the local value of the plasma $\beta$ parameter aims to capture the effects of sub-grid electron physics that are not resolved in GRMHD simulations (see \citealt{2015ApJ...799....1C}). The accretion flows in the simulations are modeled as collisionless plasmas, where the ratio of ion temperature to electron temperature ($R\equiv T_i/T_e$) is given by
\begin{equation}
R = R_{\rm{high}}\frac{\beta^2}{1+\beta^2} + \frac{1}{1+\beta^2}. 
\end{equation}
Here, $\beta\equiv P_{\rm{gas}}/P_{\rm{mag}}$ is the ratio of gas pressure ($P_{\rm{gas}}$) to the magnetic pressure ($P_{\rm{mag}}$). This prescription aims to simulate the effects of sub-grid electron heating, using this simple, physics-motivated parametric form (\citealt{2016A&A...586A..38M}, and \citealt{2019ApJ...875L...5E}). 

General relativistic ray tracing and radiative transfer is performed on the simulated system at the wavelength of $1.3\ \rm{mm}$ in order to generate the images of the black hole using the different snapshots of the simulations. For radiative transfer, the black hole mass ($M_{\rm{BH}}$) and the overall electron number density scale in the accretion disk ($n_{\rm{e}}$) set a scale for the mass and the optical depth in the accretion disk, which determines the mass accretion rate and the flux of emission in the source. It follows from the scaling properties of the emissivity of the accretion disk at 1.3 mm and from the natural scaling of the GRMHD simulations (see \citealt{2015ApJ...799....1C}) that $n_{\rm{e}}$ and $M_{\rm{BH}}$ are degenerate quantities with the resulting images and, therefore, the various interferometric observables, depending only on the product $M_{\rm BH}n_e^2$ (see a derivation and discussion in Appendix \ref{sec:appendix 1}).

We use two sets of GRMHD models for studying closure phase variability: 

\noindent (\textit{i}) The first set of models (henceforth Set A), is aimed at understanding the dependence of the closure-phase variability on the properties of the accretion flow, i.e., MAD vs.\ SANE magnetic field configuration, the plasma prescription, and the electron number density scale ($n_{\rm{e}}$). In particular, they include SANE and MAD models for a single spin, three different values of $R_{\rm{high}}$, and five different values of $n_{\rm{e}}$ (defined at a fiducial mass of $M_{\rm{BH}} = 6.5 \times 10^9 M_\odot$). For this value of the black-hole mass, this range allows us to explore compact fluxes that are up to a factor of a few above and below the nominal value of $0.5\ \rm{Jy}$. Each model is used to generate 1024 images with a time cadence of 10 $GM/c^3$, amounting to a total of $\sim 30000$ images in this set. Ray tracing and radiative transfer calculations for this simulation set is carried out using \texttt{GRay}. 

\noindent (\textit{ii}) The second set of models (henceforth Set B) explore a comprehensive set of black hole spins and $R_{\rm{high}}$ of the accretion plasma, which are used in \citet{2019ApJ...875L...5E} in the interpretation of the EHT observations. The electron number density scale ($n_{\rm{e}}$) in this simulation set is tuned {\em a priori\/} such that the total flux of radiation at $1.3\ \rm{mm}$ is equal to $0.5\ \rm{Jy}$ for a fiducial mass of $M_{\rm{BH}} = 6.2 \times 10^9 M_\odot$. Each model in this simulation set is used to generate  600-1000 images at a time cadence of 5 $GM/c^3$, amounting to $\sim 50000$ images in this set. Ray tracing and radiative transfer calculations for this simulation set is carried out using \texttt{ipole}.

The ray tracing and radiative transfer tracing calculations on the GRMHD simulations performed using \texttt{ipole} and \texttt{GRay} ignore the effects of finite light travel time. \citet{2018A&A...613A...2B} showed that the affect of the approximation on the lightcuve of flux density in a simulation is restricted to only a few percent. Depending upon the nature of spatial correlations in structural variability in the simulation, the computed images may carry either time-correlated or time-uncorrelated features. These features will have an effect on the amplitude of variability inferred. In this study, we ignore these effects in inferring the variability in the images.

The parameter space studied in sets A and B are listed in Table \ref{tab:simulation library}. 
Although the mass of the black hole $M_{\rm{BH}}$ does not enter as an independent parameter in the calculation of the black-hole images, it affects our interpretation of the observations in the following ways:

(\textit{i}) The characteristic time unit ($\tau$) in the GRMHD simulations is set by the mass of the black hole as $\tau = GM_{\rm BH}/c^3$. A higher mass of the black hole implies that an observation span of 6 days would translate to a shorter elapsed time in units of $\tau$. 

(\textit{ii}) The angular size of the image as seen by a distant observer depends on the black-hole mass ($\Delta\theta\sim\lambda/M_{\rm BH}$), which implies that the EHT probes different regions in the Fourier space for black holes of different masses. A higher mass of the black hole translates to the EHT probing structures at smaller length scales. 

The mass of M87 has been measured using two methods prior to the EHT observations (\citealt{2019ApJ...875L...5E}). \citet{2011ApJ...729..119G} measured a mass of $M_{\rm{BH}} = 6.6 \pm 0.4 \times 10^9 M_\odot$ using stellar dynamics. \citet{2013ApJ...770...86W} measured a mass of $M_{\rm{BH}} = 3.5^{+0.9}_{-0.7} \times 10^9 M_\odot$ by studying emission lines from ionized gas. The GRMHD models in simulation sets A and B give images for which the ring sizes and widths correspond to masses in the range inferred by the stellar dynamics measurement of \citet{2011ApJ...729..119G} (see \citealt{2019ApJ...875L...6E}). Because of this, we allow the black hole mass to vary in the range $5 \times 10^9 M_\odot \leq M_{\rm{BH}} \leq 9.5 \times 10^9 M_\odot$ in both sets of the simulations.

The orientation of the black-hole spin axis has been constrained by long-wavelength observations of its jet to be at a position angle of $\sim 288^\circ$ North of East and at an inclination of 17$^\circ$ with respect to the line of sight (see discussion in \citealt{2019ApJ...875L...6E}). We trace the Set A models at an inclination of $17^\circ$. In Set~B, we trace the positive spin models at an inclination of $163^\circ$ and the negative spin models at $17^\circ$. 

The simulations typically run for lengths that correspond to $\sim 10^4 GM/c^3$. Here, we only use images obtained from the relaxed turbulent state with a slowly varying mass accretion rate. This corresponds to simulation times in the range of $5\times 10^3 \leq t \leq 10^4 GM/c^3$. 

\begin{table*}[t]

\caption{GRMHD Simulation Model Parameters}
\begin{center}
\begin{threeparttable}
\noindent \begin{tabular}{ c c c }
\toprule
Parameter \tnote{a} & \multicolumn{2}{c}{Parameter Space} \\
\cmidrule{2-3}
& Set A & Set B \\
\midrule
Model Type  & MAD, SANE & MAD, SANE \\
Black Hole Spin \tnote{b} & $+0.9$ &  $0.0$, $\pm0.5$, $\pm0.94$ \\
Plasma $R_{\rm{high}}$ & $1$, $20$, $80$ & $1$, $10$, $20$, $40$, $80$, $160$ \\
Electron Number Density ($n_{\rm{e}}$) \tnote{c} & $1,2.5,5,7.5,10$ ($\times 10^5 \text{cm}^{-3}$) & tuned to constant flux \\
Time Cadence & $10\ GM/c^3$ & $5\ GM/c^3$ \\
\bottomrule
\end{tabular}
\begin{tablenotes}
\item Notes
\item[a] The $u-v$ coordinates of the baselines are rotated to align the spin axis of the black hole at a position angle of $288^\circ$ East of North.
\item[b] The positive spin models are ray traced at an inclination of $163^\circ$ and negative spin models at $17^\circ$ in set A. Set B models are ray traced at an inclination of $17^\circ$.
\item[c] The electron number density is defined at a fiducial mass of $6.5 \times 10^9 M_\odot$ for set A and $6.2 \times 10^9 M_\odot$ for set B. 
\end{tablenotes}
\end{threeparttable}
\end{center}
\label{tab:simulation library}
\end{table*}

\section{Closure Phase Variability in GRMHD Simulations}
\label{sec:sec4}

In order to analyze visibility phase and closure phase variability in GRMHD models, we first transform the GRMHD snapshots into the visibility space by performing a two-dimensional Fourier transform on each of the images. We then compute the amplitude and phase from the complex visibilities at each location in the $u-v$ plane. 

Since the visibility phase is a directional quantity, we need to employ directional statistics to infer the standard deviation in a time-series. For a given time-series of $n$ phases $\theta_i$, we compute its directional dispersion as
\begin{equation}
\label{eq:directional dispersion}
D = \frac{1}{n}\sum_{i=1}^n \left\{1-\cos(\theta_i - \bar\theta)\right\}, 
\end{equation}
where $\bar\theta$ represents the circular mean, defined as 
\begin{equation}
\bar\theta = \tan^{-1}\left\{\frac{\sum_{i=1}^n\sin(\theta_i)}{\sum_{i=1}^n\cos(\theta_i)}\right\}.
\end{equation}
In the limit of small deviations in the time-series from the circular mean $\bar\theta$, the directional dispersion can be approximated as
\begin{equation}
\label{eq:small angle D}
D \approx \frac{\sigma^2}{2}, 
\end{equation}
where $\sigma$ is inferred as a standard deviation of the time-series $\theta_i$ for small fluctuations about $\bar\theta$.

\begin{figure*}[t]
	\centerline{
	\includegraphics[ width = 0.5\textwidth]{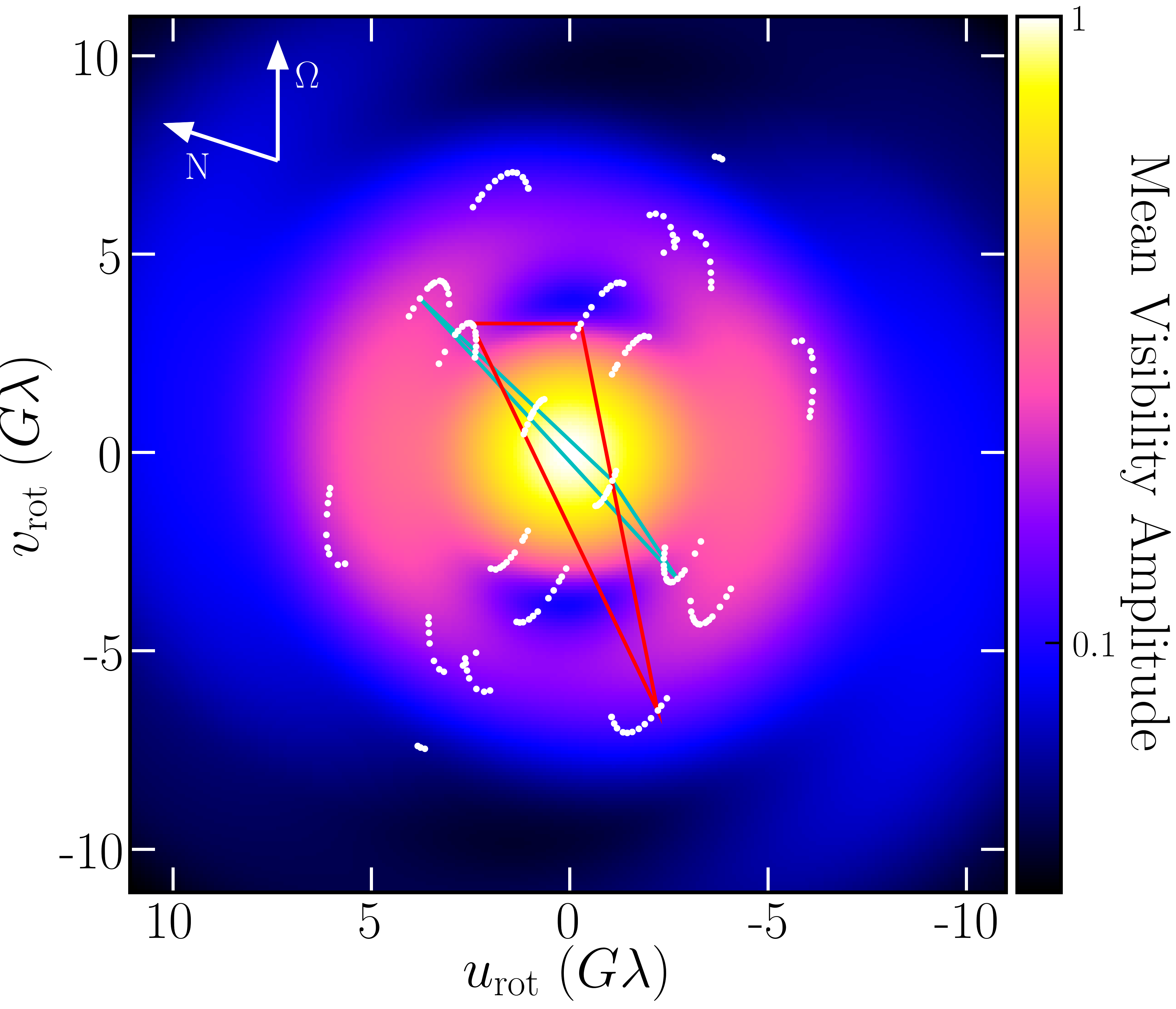}
	\includegraphics[ width = 0.5\textwidth]{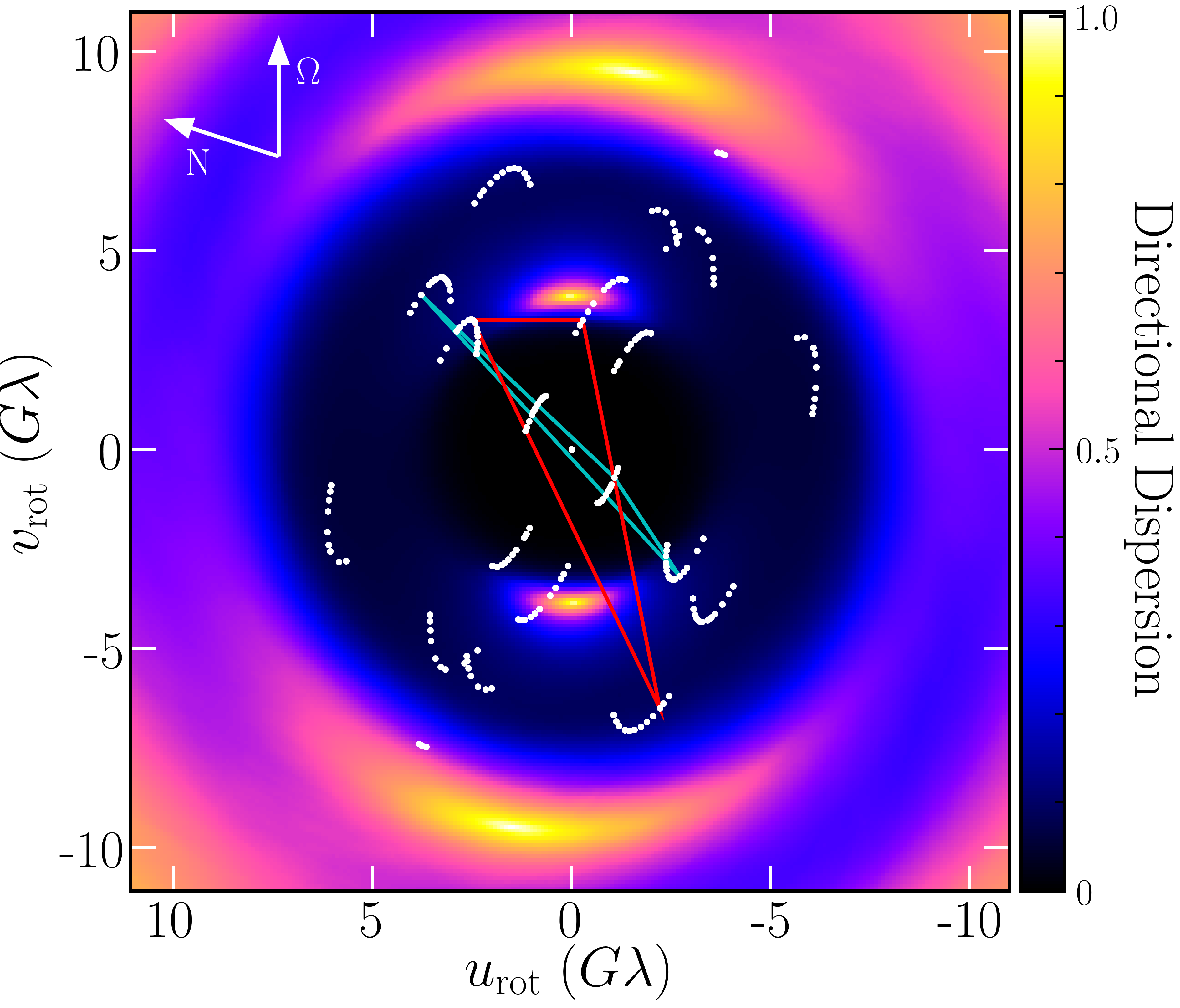}
}	
\caption{Normalised mean visibility amplitude map on a logarithmic scale (\textit{left}) and directional dispersion map (see Equation \ref{eq:directional dispersion}) of visibility phases (\textit{right}) for the SANE, $a = +0.5$, $R_{\rm{high}} = 1$ model from simulation set B. The black hole spin points upwards as indicated by the white arrow labeled as $\Omega$. The EHT baseline coverage (on April 11) are shown by the white dots, and are rotated such that the spin points to a position angle of $288^\circ$ east of north (with north indicated by the white arrow labeled as N and the $u_{\rm{rot}}-v_{\rm{rot}}$ plane representing the rotated $u-v$ plane).  The cyan and the red triangles indicate the ALMA-LMT-SMT (at UTC 3:32:5.0003 h / GMST 16:26:37.1531 h) and ALMA-SMA-LMT (at UTC 5:00:5.0001 h / GMST 17:54:51.6089 h) triangles respectively. The second triangle includes the SMA-LMT baseline that crosses a visibility minimum, which translates to a high level of variability in closure phase introduced by structural changes in the image. The mass of the black hole is set to $M_{\rm{BH}} = 7.5 \times 10^9 \ M_{\odot} $ }
\label{fig:sample heatmap}
\end{figure*}

We use this quantity to construct heat maps of phase variability in order to understand the dependence of variability on the $u-v$ coordinates (see also \citealt{2017ApJ...844...35M}). We present an example of such a heat map for one of the simulations in set B in Figure~\ref{fig:sample heatmap}. The left panel shows the mean visibility amplitude on the $u-v$ plane computed across the simulation run and reveals deep visibility minima that are aligned with the spin axis of the black hole. The right panel shows a heat map of visibility phase dispersion. There are "hot regions" of visibility phase variability, i.e., localized regions in $u-v$ space that show large dispersions. These regions lie primarily along the spin axis of the black hole and at baseline lengths that correspond to the deep minima in visibility amplitudes. In fact, a natural anticorrelation between phase variability and mean visibility amplitude is observed in all the GRMHD models, which was also explored in \citet{2017ApJ...844...35M} (see their Figures 3 and 4). 

Based on this understanding, one expects the closure phase variability computed along various triangles to depend on the locations of the three vertices in the $u-v$ space. In particular, the localized nature of variability implies that closure triangles that have baselines crossing the hot regions are expected to exhibit high variability in closure phases, whereas triangles with vertices in quiet parts of the $u-v$ space should show a low level of variability. In Figure~\ref{fig:sample heatmap}, we show examples of two such triangles: the ALMA-SMA-LMT triangle has a baseline on a hot region and is expected to be highly variable, while a low-variability closure triangle (ALMA-LMT-SMT) avoids the hot regions. 

The azimuthal extent of the hot regions of phase dispersion in the $u-v$ plane depends on the degree of symmetry of the underlying image. This is shown in Figure~\ref{fig:ne_heatmaps}, where each column corresponds to a simulation from set A with a different electron density scale $n_e$, and all other black hole and plasma parameters held fixed. Qualitatively, the two effects of increasing the electron density scale are a higher level of symmetry in the bright emission ring and an increase in its fractional width. The more symmetric images lead to a larger azimuthal extent in the minima in the visibility amplitudes (middle panels) and in the hot regions of phase dispersion (lower panels). The increased fractional width of the images lead to the locations of both the visibility amplitude minima and the hot regions in phase dispersion to appear at smaller baseline lengths.

\subsection{Closure Phase Variability and Compliance Fractions}

The phase dispersions discussed in the previous section were calculated across the entire span of the simulations, which corresponds to many dynamical timescales. In order to compare the simulations to the observed variability in M87, however, we need to calculate the degree of variability for a timespan of $\Delta t =6$~days, which is the longest separation between the four observations in 2017. Furthermore, because observations only yield closure phases, hereafter we focus on this quantity to facilitate comparisons. 

We define the variance in closure phase $\phi_{\rm cp}(t)$ for a time-span of $\Delta t = 6\ \rm{days}$ using Equation~(\ref{eq:small angle D}) as
\begin{equation}
\label{eq:continuous D}
\sigma^2 (t_0) = \frac{1}{\Delta t}\int_{t_0}^{t_0 + \Delta t}dt \ 2\left[1-\cos\{\phi_{\rm cp}(t) - \bar\phi_{\rm cp}\}\right].
\end{equation}

In order to reduce the discretization noise when calculating this quantity, we perform a linear interpolation between simulation time-steps and construct a continuous function $\phi_{\rm cp}(t)$ representing the evolution of closure phases. We use fixed $u-v$ locations taken at a median time over the night of observation, in order to construct this function. This ensures that we separate the evolution of closure phases due to the changing locations of the baselines from the structural variability of the images that we are interested in inferring.

\begin{figure}
	\centerline{
	\includegraphics[width = 0.5\textwidth]{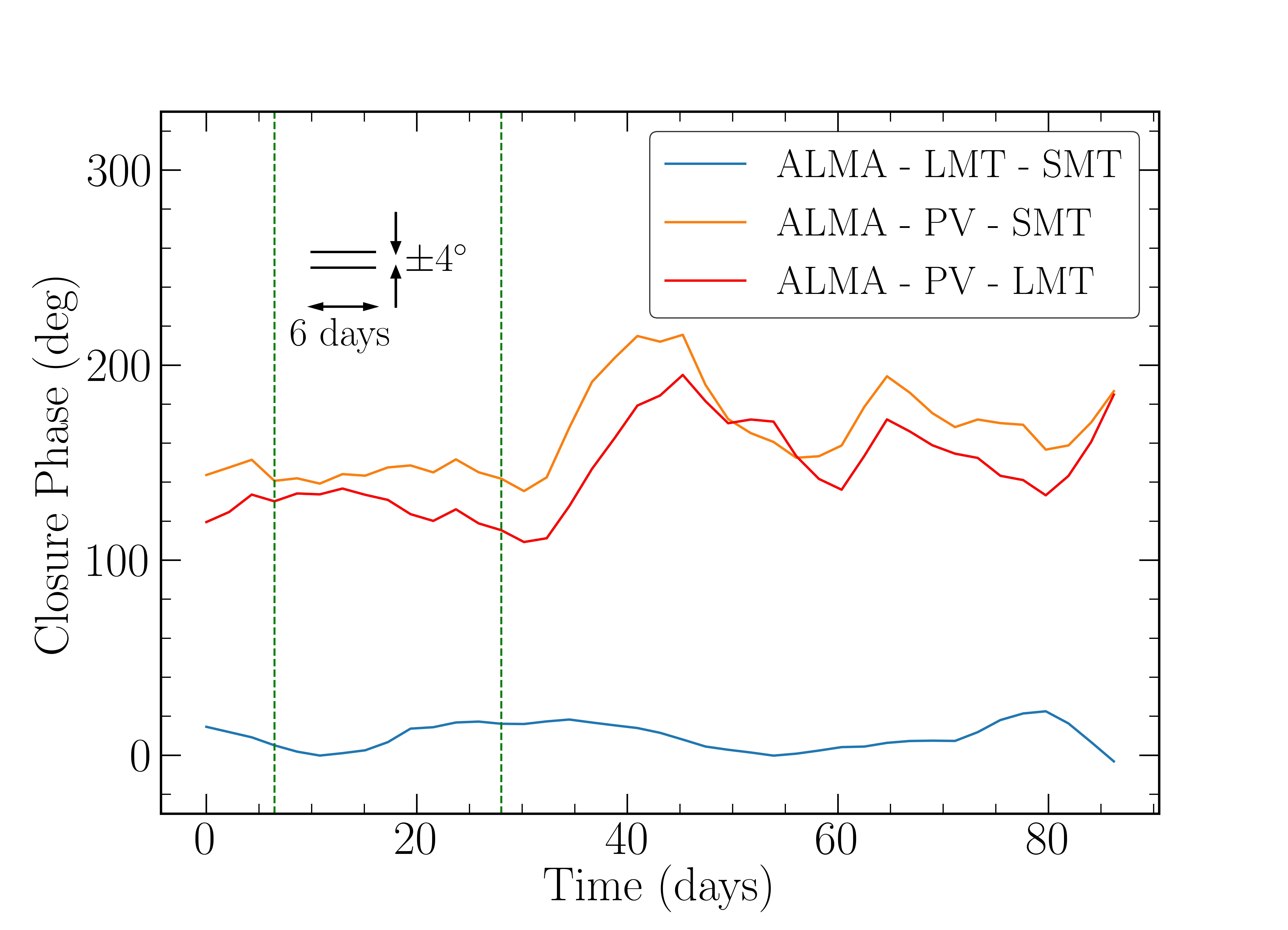}
	}
    \caption{Closure phases for three triangles plotted as a function of time for a particular segment of the same simulation (set B - MAD $a = +0.94$, $R_{\rm{high}}=20$). The black parallel lines indicate the time scale of 6 days and a standard deviation of $\sim4^\circ$. The section within the green dashed vertical lines indicates a region of low-variability, and hence counts towards the compliance fraction of the model. }
\label{fig:cp_variability}
\end{figure}

In Figure~\ref{fig:cp_variability}, we show the evolution of closure phase with time during one segment of a simulation chosen from set B, that has a MAD configuration with parameters $a = +0.94$ and $R_{\rm{high}}=20$. As is evident from this evolution, there are time spans where the closure phase shows little variability (e.g., the times between the green vertical lines) and others where there are substantial swings of order 
$\pi$ over a short period of time that is comparable to the length of the observations. Given the low degree of variability observed for some triangles in M87 data (see $\S$ \ref{sec:section 2}), this motivates us to define a metric that quantifies how common such periods are in a given model. 

\begin{figure*}
	\centerline{
	\includegraphics[width = 0.5\textwidth]{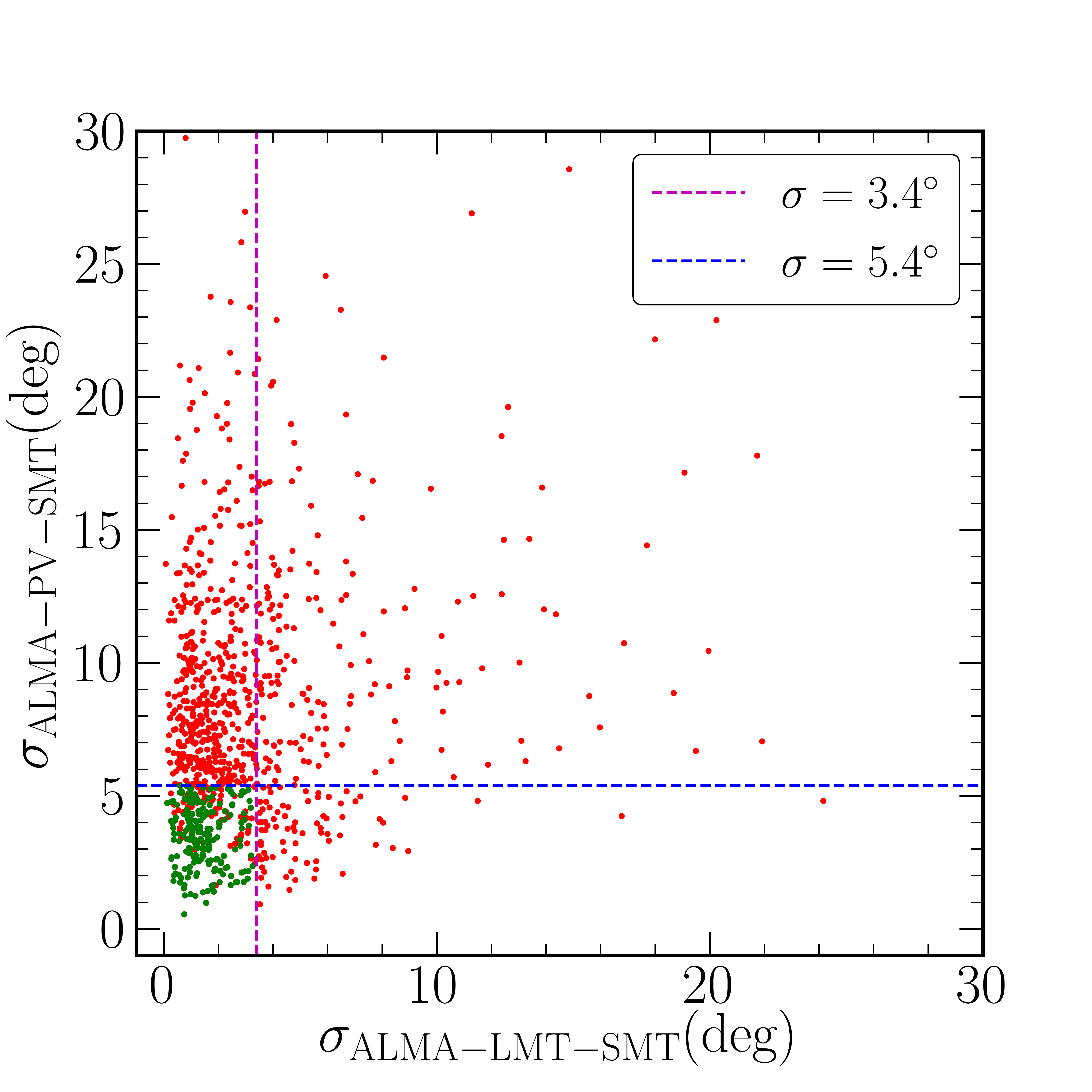}
	\includegraphics[width = 0.5\textwidth]{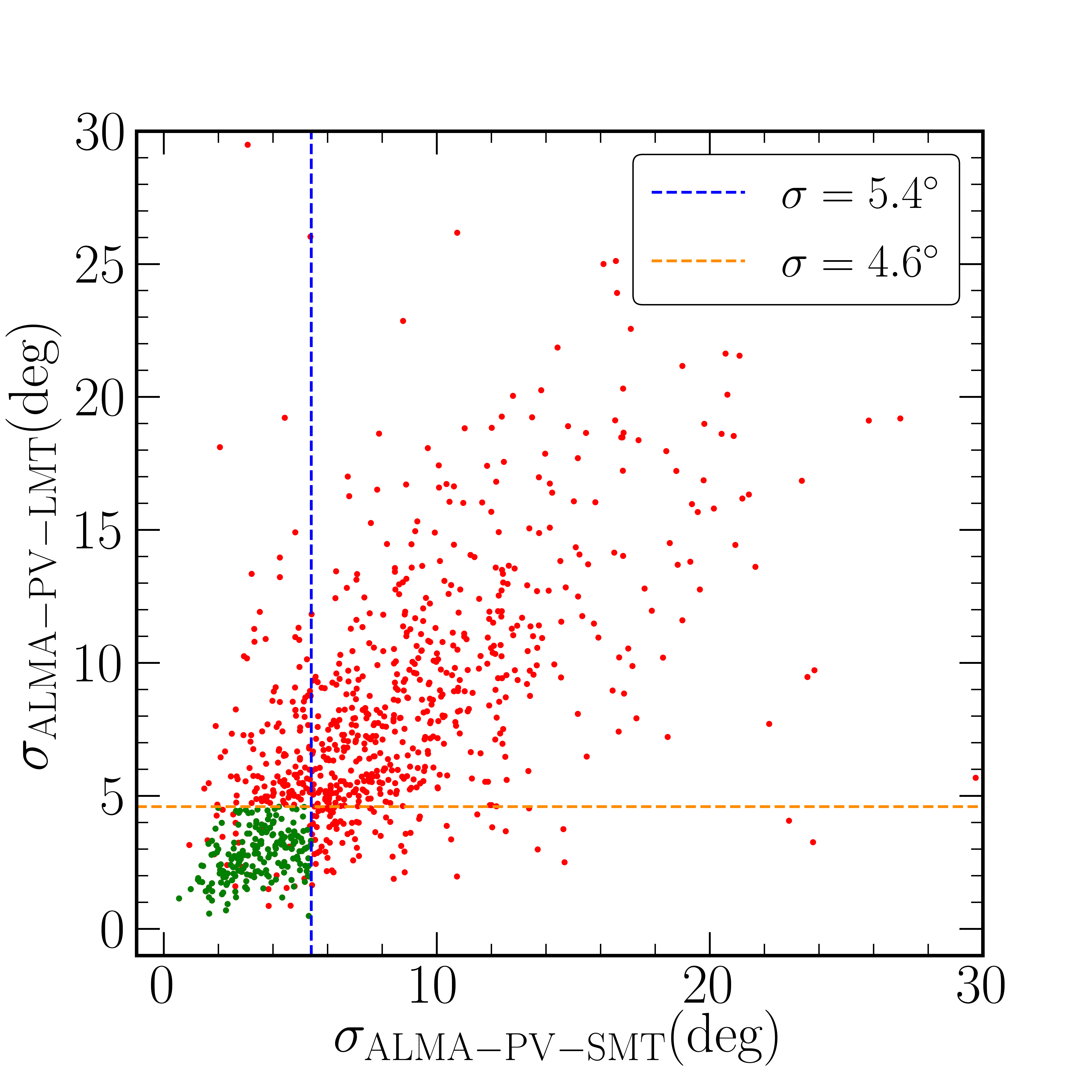}
				}	
\caption{Two dimensional scatter plots of inferred standard deviation in closure phases in three triangles across one simulation (simulation set B - MAD $a = +0.94$, $R_{\rm{high}}=20$). The horizontal and vertical dashed lines indicate the observational bounds for the standard deviation for each triangle, as described in $\S$ \ref{sec:section 2}.}
\label{fig:compliance_computation}
\end{figure*}

\begin{figure*}
	\centerline{\includegraphics[width=0.33\textwidth]{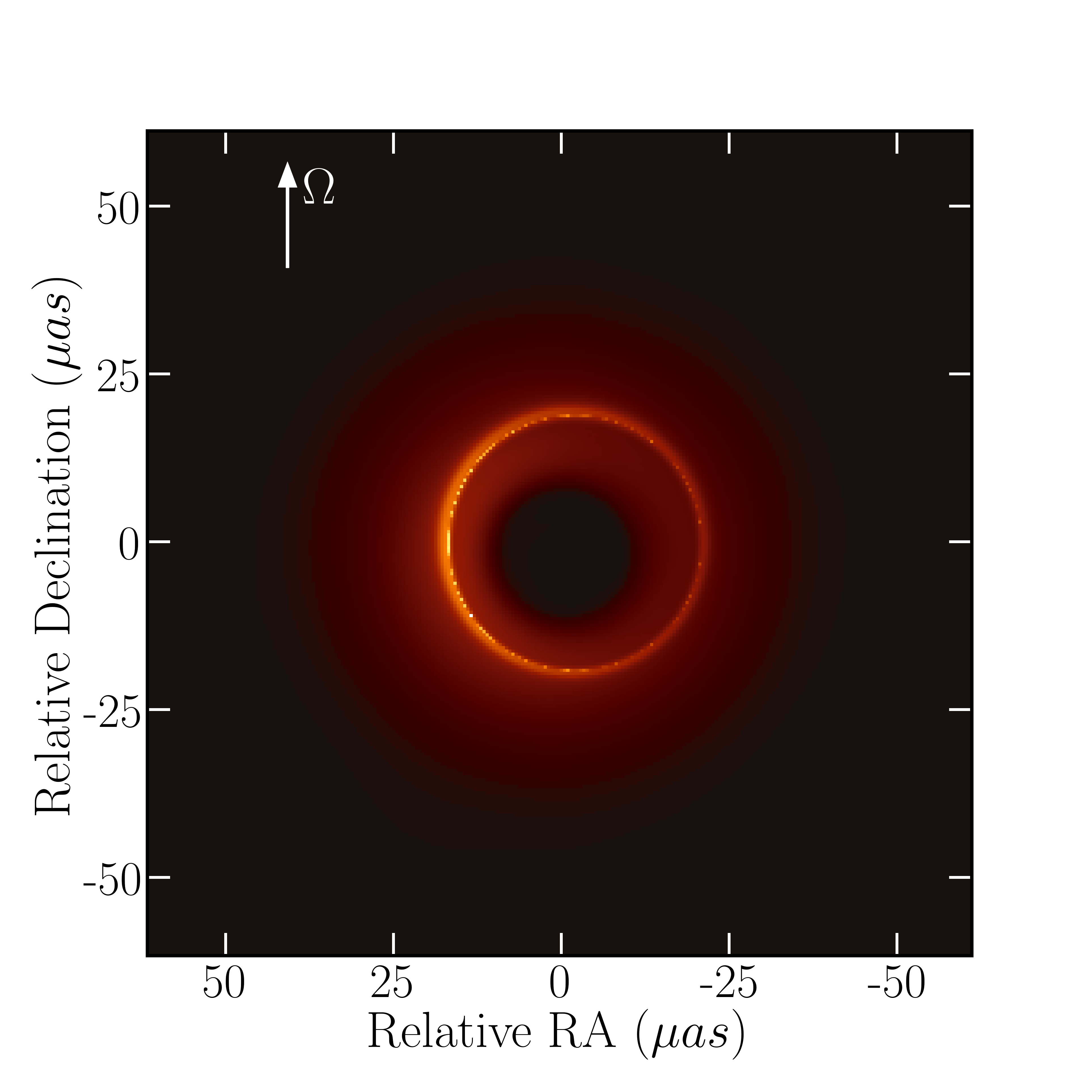}
				\includegraphics[width=0.33\textwidth]{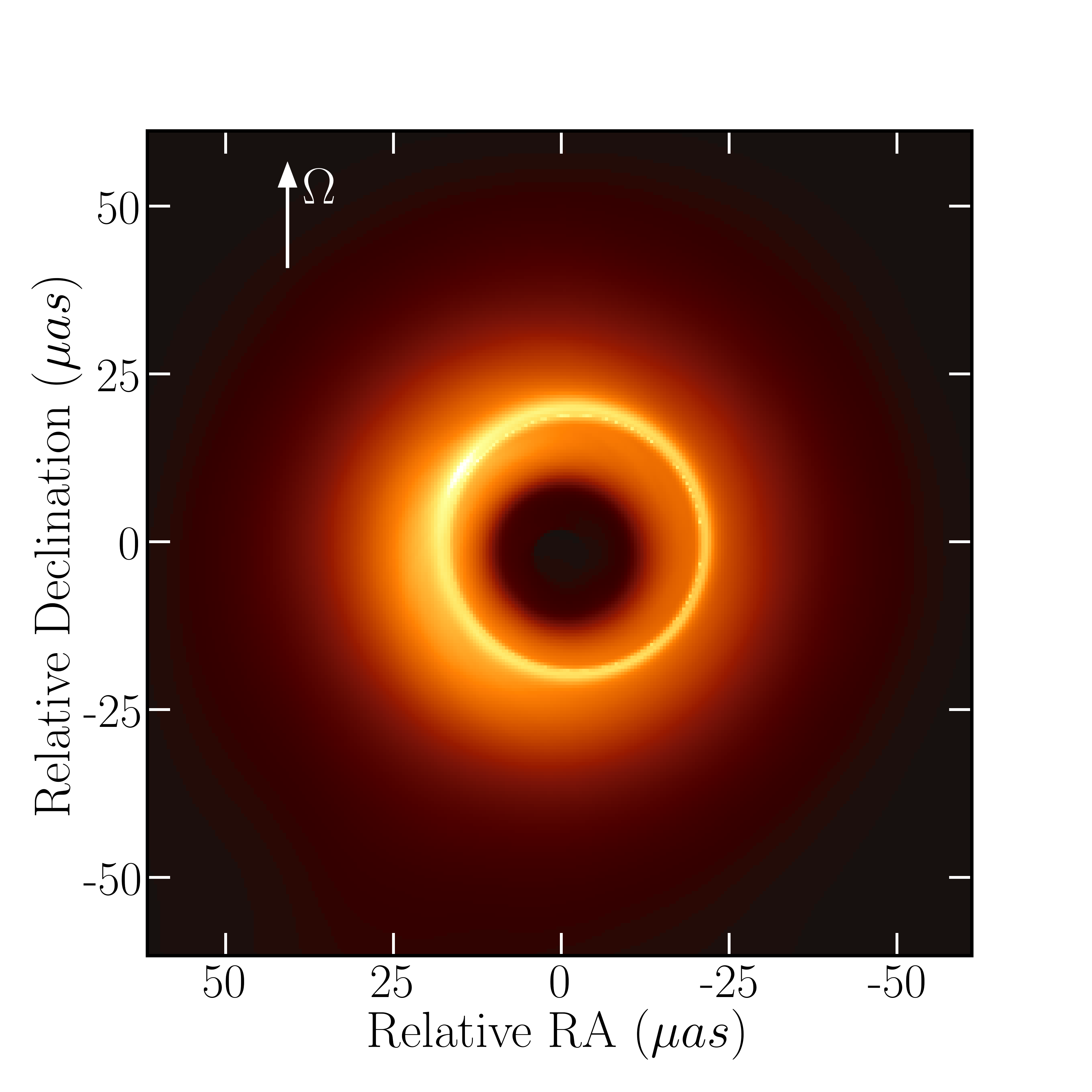}
				\includegraphics[width=0.33\textwidth]{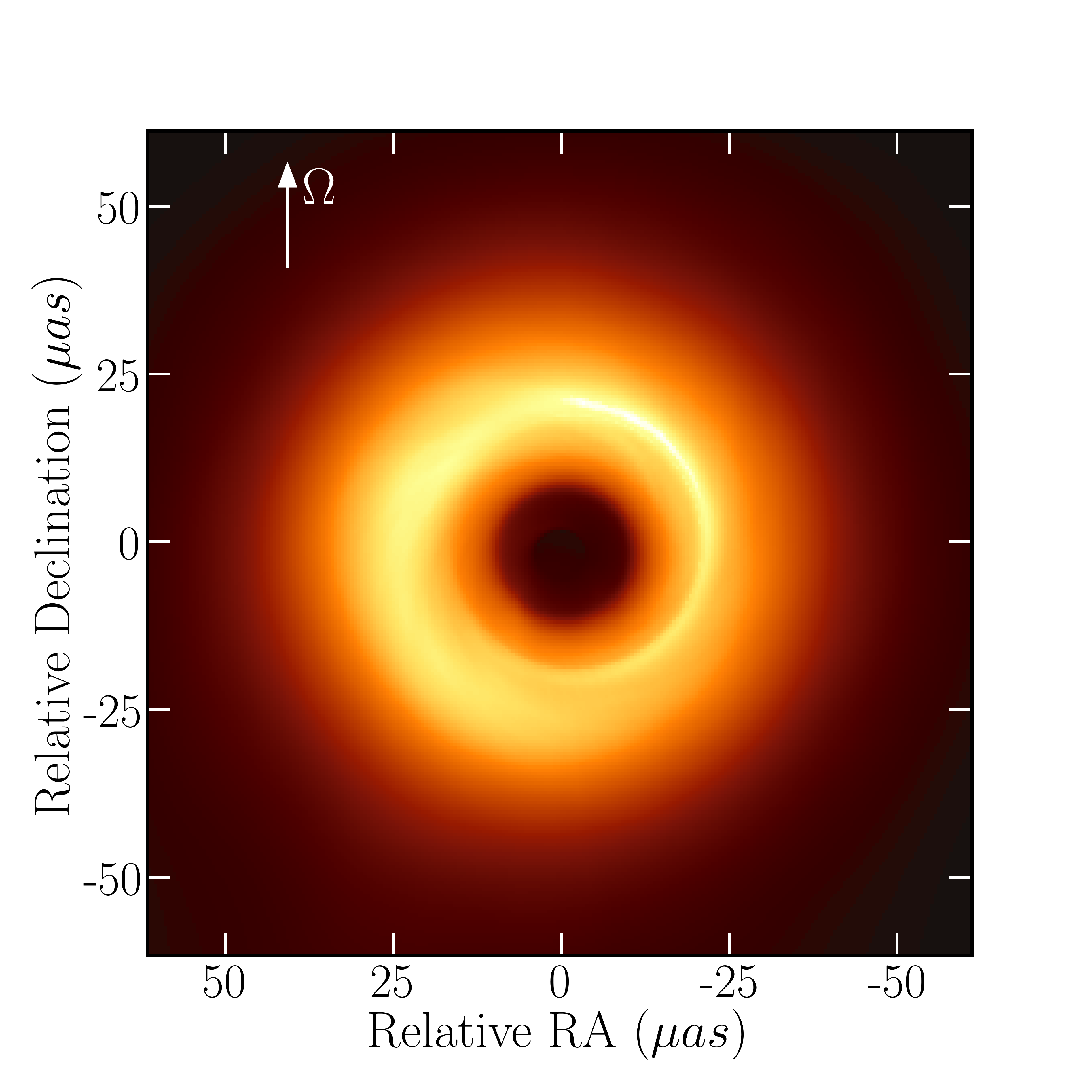}}
	\centerline{\includegraphics[width=0.33\textwidth]{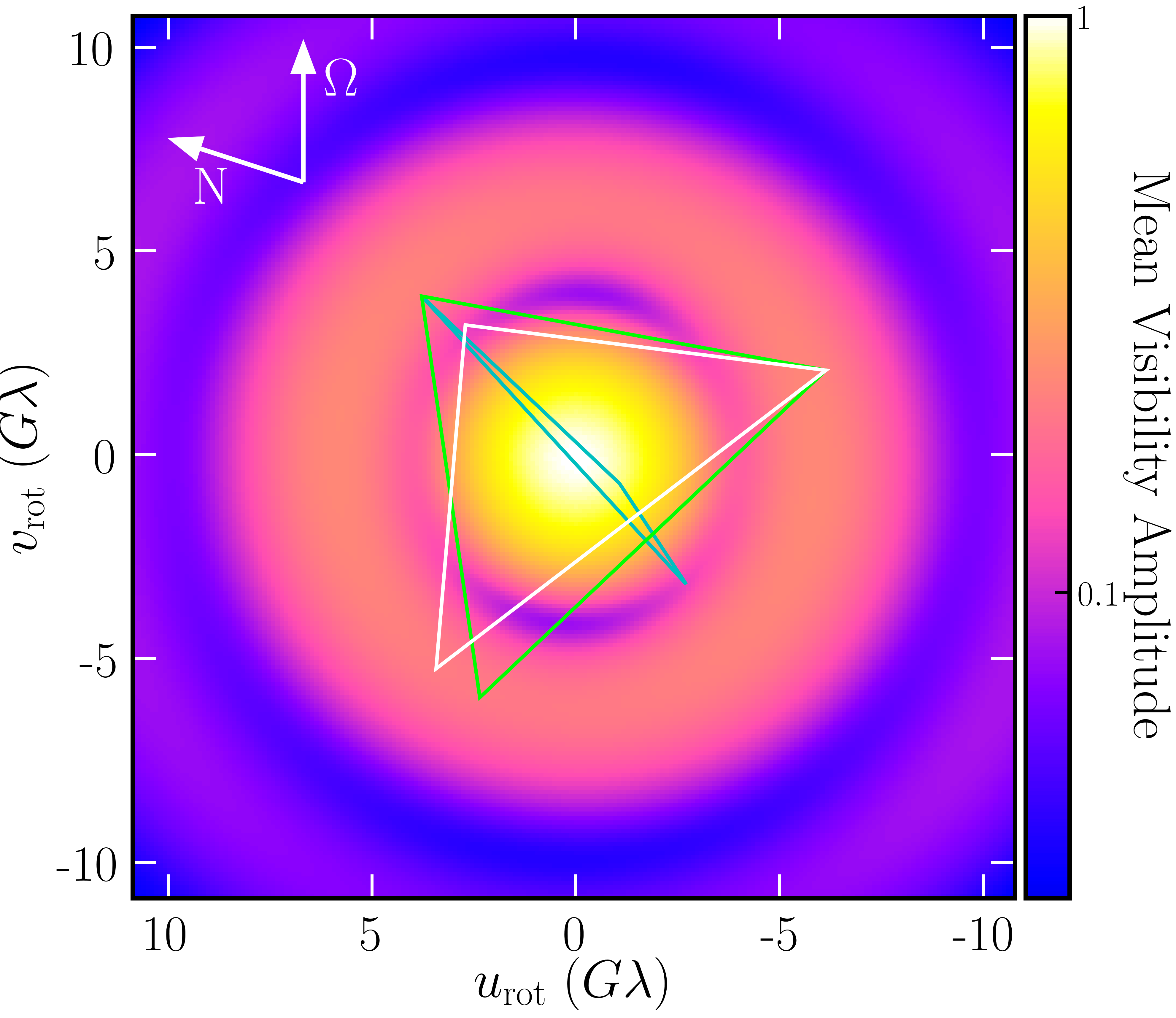}
				\includegraphics[width=0.33\textwidth]{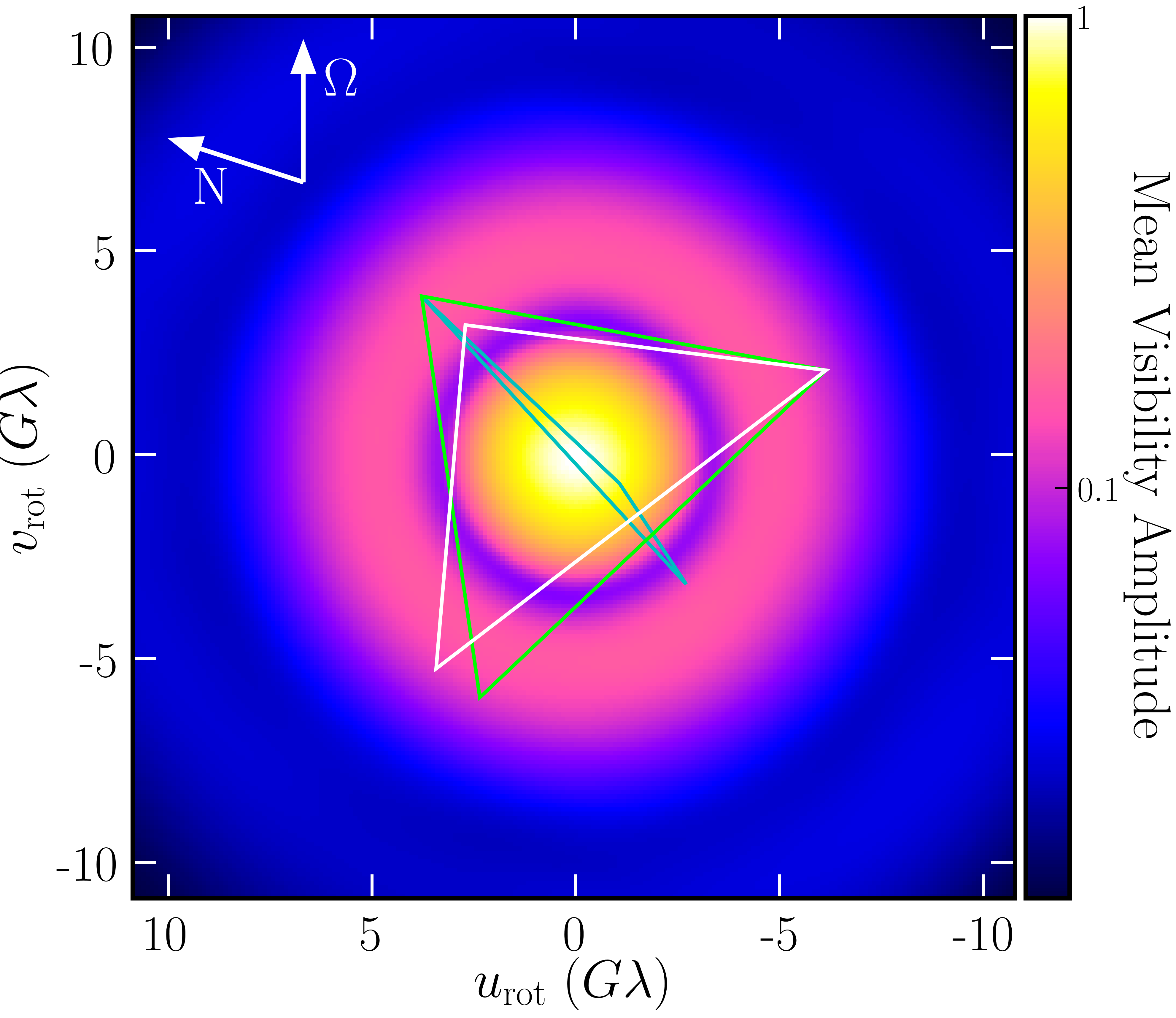}
				\includegraphics[width=0.33\textwidth]{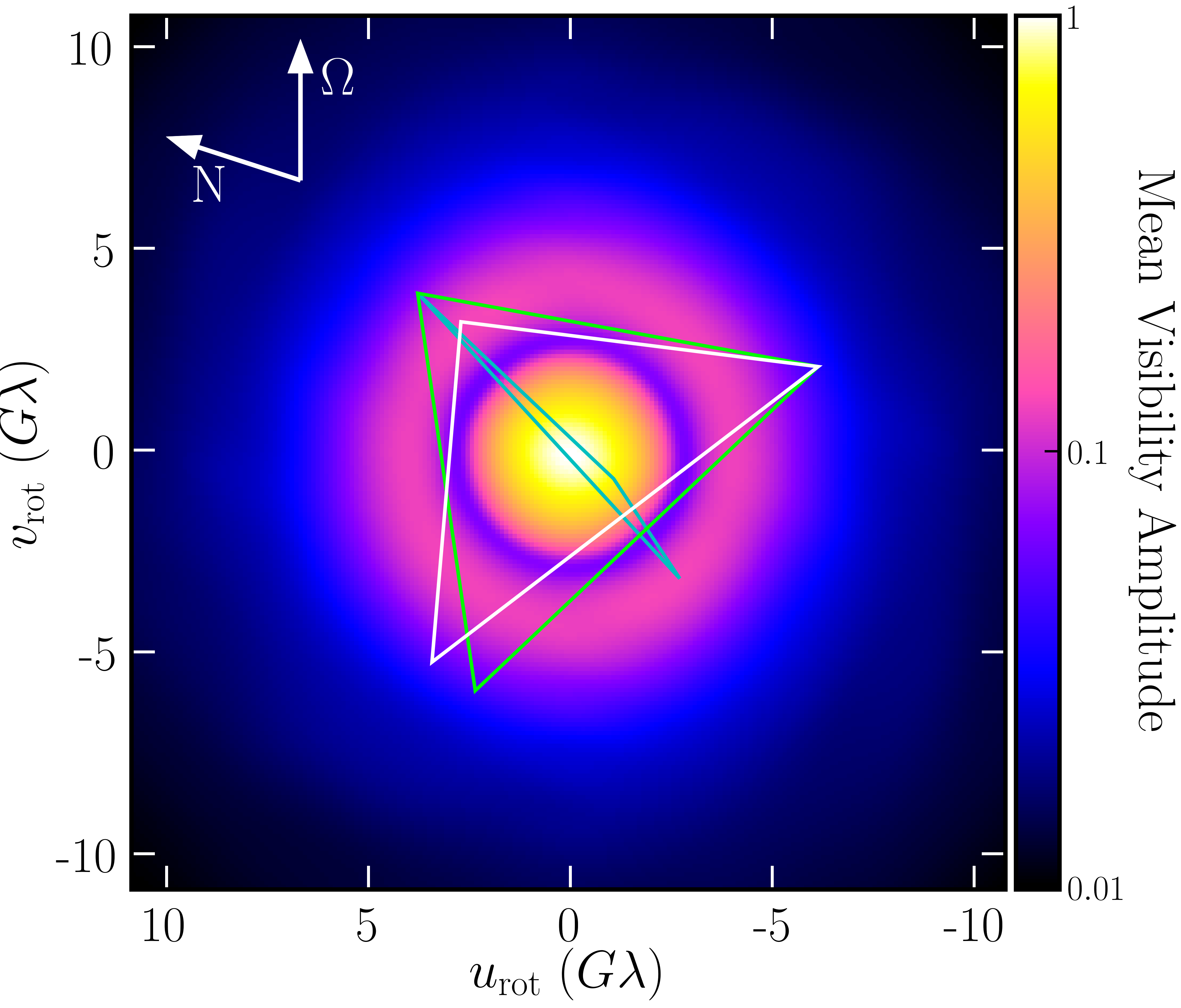}
				}
	\centerline{\includegraphics[width=0.33\textwidth]{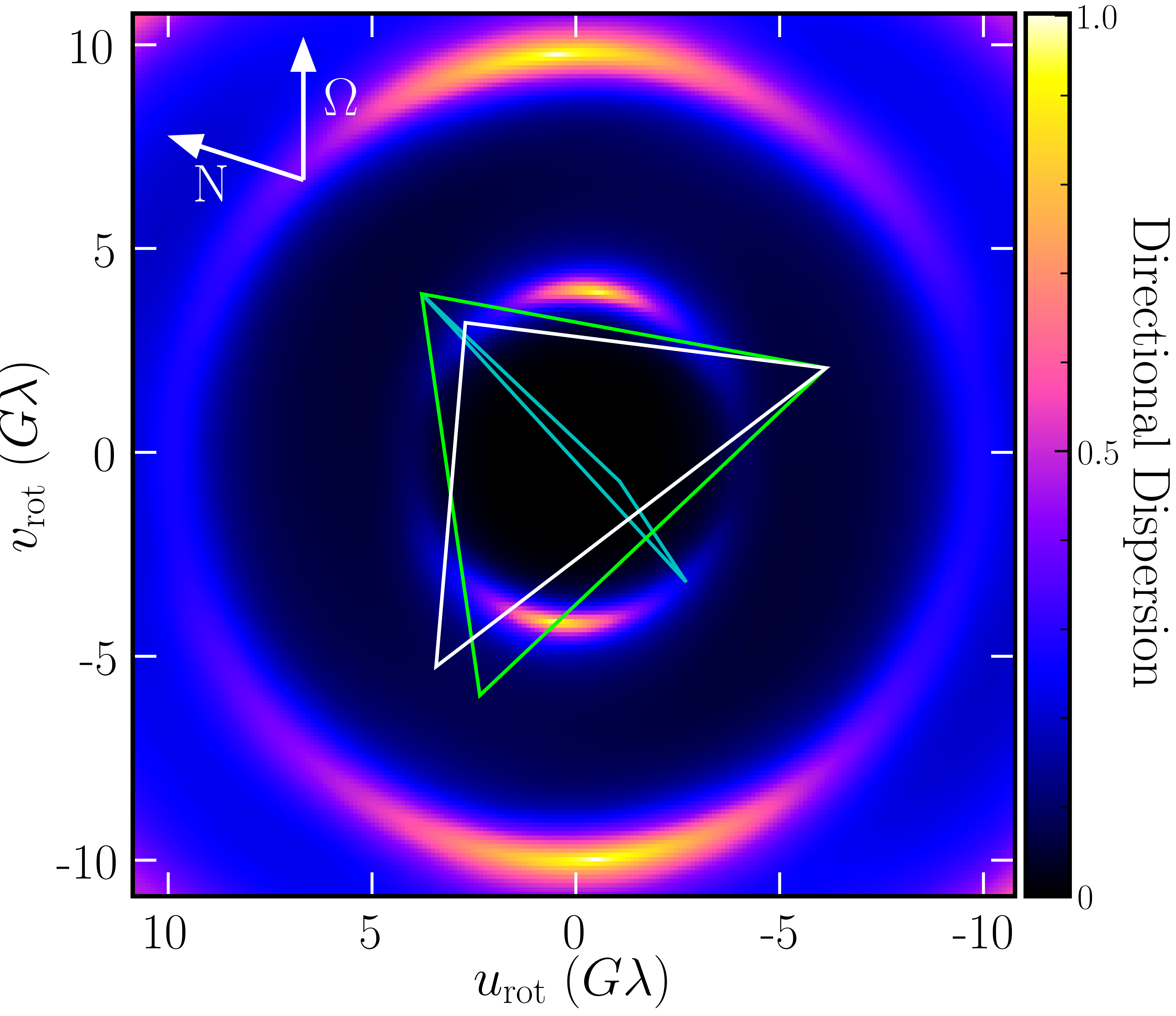}
				\includegraphics[width=0.33\textwidth]{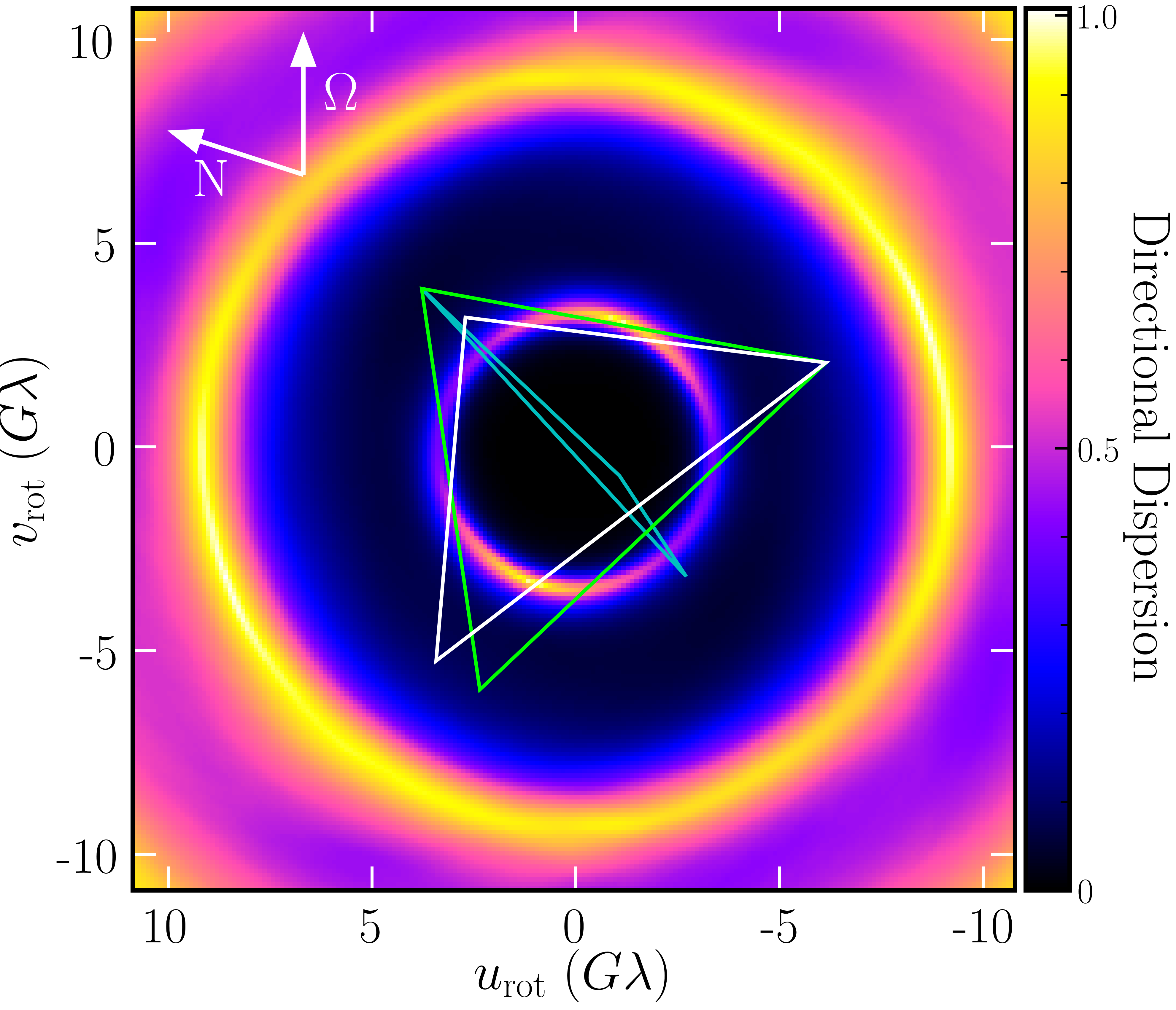}
				\includegraphics[width=0.33\textwidth]{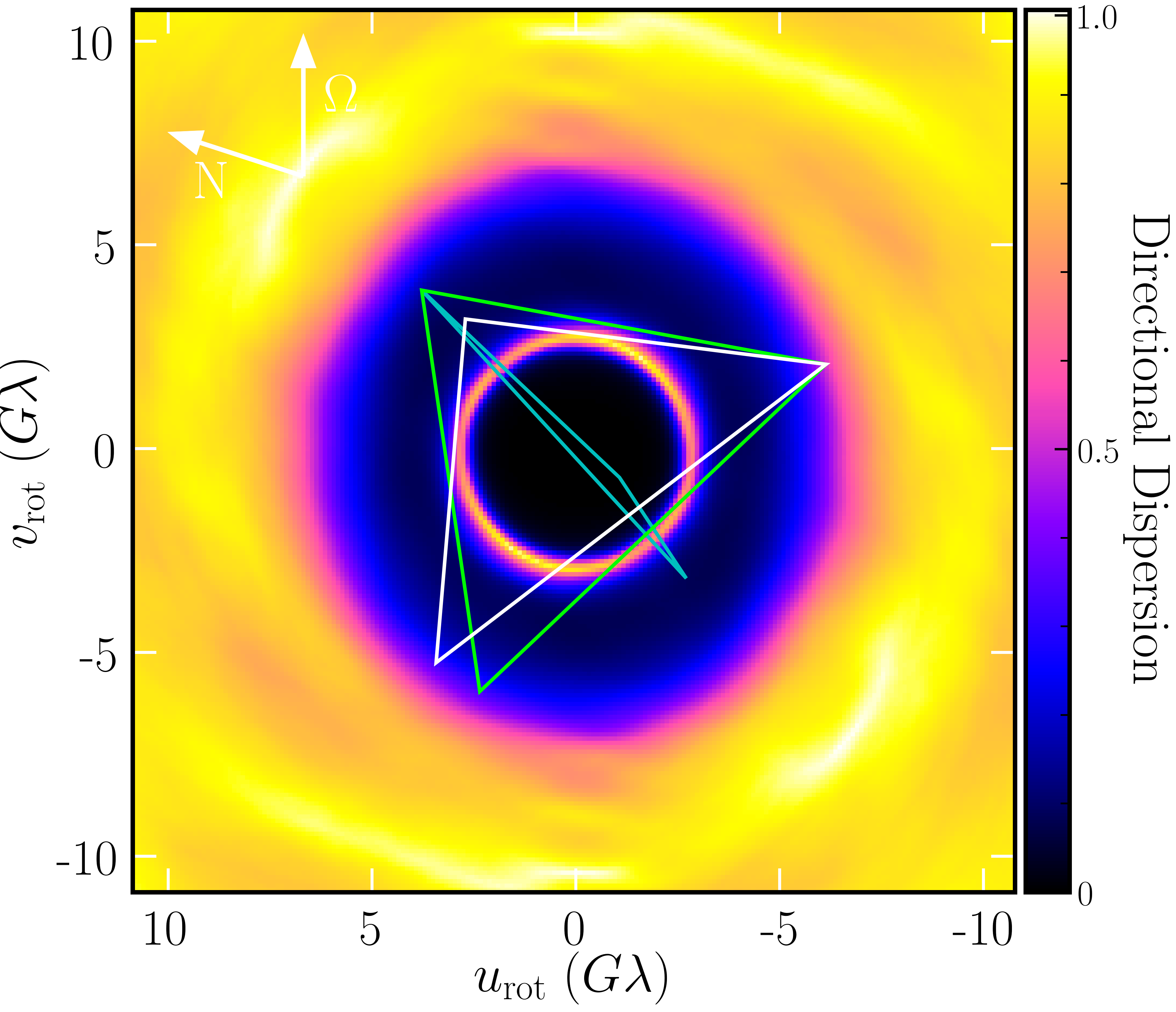}}
\caption{The average images (\textit{top}), normalized average visibility amplitudes (\textit{middle}), and directional phase dispersion (\textit{bottom}) of the MAD, a = 0.9, $R_{high} = 20$ simulation. Models with three electron number densities - $n_e = 1 \times 10^5\ \text{cm}^{-3}$ (\textit{left}), $n_e = 5 \times 10^5\ \text{cm}^{-3}$ (\textit{middle}), and $n_e = 1 \times 10^6 \text{cm}^{-3}$ (\textit{right}), are mapped in this figure. The white arrows labeled as $\Omega$ indicate the spin vector of the black hole, and the ones labeled as N indicate the north direction. The coordinates $(u_{\rm{rot}},v_{\rm{rot}})$ represent the $u-v$ coordinates rotated such that the spin points to a position angle of $288^\circ$ east of north. It is evident that a model with lower $n_e$ exhibits emission dominant from the photon ring, and has a lower variability in visibility phases outside the first visibility minimum. The three triangles plane indicate the low-variability triangles - ALMA-LMT-SMT (\textit{cyan}), ALMA-PV-SMT (\textit{green}), and ALMA-PV-LMT (\textit{white}) at UTC 3:32:5.0003 h/ GMST 16:26:37.1531 h. The colormap in visibility amplitude is in logarithmic scale with the maximum value is normalised to unity. }
\label{fig:ne_heatmaps}
\end{figure*}

Figure~\ref{fig:compliance_computation} shows the standard deviation in closure phase variability computed from all 6-day segments in the simulation shown in Figure~\ref{fig:cp_variability}, for the three triangles that have observationally shown a small degree of variability. The dashed lines correspond to thresholds $\sigma_1, \sigma_2, \sigma_3$ of this observed variability. In Figure \ref{fig:cp_variability}, we choose as the most likely values of $\sigma_{c_0}$ in the corresponding triangles as the thresholds (see Table \ref{tab:closure triangles}). The green points indicate the occurrences of 6-day segments that are consistent with observations. We define the quantity $\mathscr{F}(\sigma_1,\sigma_2,\sigma_3)$ of a model as the fraction of 6-day segments that show a level of variability consistent with the observed one; i.e., the ratio of green points to the total number of points in this figure. 

Formally, $\mathscr{F}(\sigma_1,\sigma_2,\sigma_3)$ can be written as 
\begin{equation}
\label{eq:defn F}
\begin{split}
\mathscr{F}(\sigma_1,\sigma_2,\sigma_3) = \int_0^{\infty}d\sigma'_1\int_0^{\infty}d\sigma'_2\int_0^{\infty}d\sigma'_3\ P(\sigma_{\rm{sys}})\\
\times P_{\rm{model}}(\sigma'_1,\sigma'_2,\sigma'_3 | \sigma_{\rm{sys}}),
\end{split} 
\end{equation}
where $P(\sigma_{\rm{sys}})$ is the prior in the systematic uncertainty $\sigma_{\rm{sys}}$ and $P_{\rm{model}}(\sigma'_1,\sigma'_2,\sigma'_3|\sigma_{\rm{sys}})$ is the three-dimensional distribution of circular standard deviations $\sigma'_1,\sigma'_2,\sigma'_3$ computed in the three low-variability triangles using Equation \ref{eq:continuous D}. The observed variability, as quantified in $\S$ \ref{sec:section 2}, is a combination of the true intrinsic variability of the source and the systematic errors in measurement. Because of our limited knowledge of the systematic uncertainties and their priors, we assume a form such that equation (13) can be written as

\begin{equation}
\label{eq:defn F prior}
\mathscr{F}(\sigma_1,\sigma_2,\sigma_3) = \int_0^{\sigma_1}d\sigma'_1\int_0^{\sigma_2}d\sigma'_2\int_0^{\sigma_3}d\sigma'_3\ P_{\rm{model}}(\sigma'_1,\sigma'_2,\sigma'_3), 
\end{equation}
i.e., simply assuming that the combined intrinsic degree of variability and the systematic uncertainties cannot exceed the inferred values.

Because our inferred degree of variability from observations itself has formal errors associated with it, as described by the marginalized posteriors $P(\sigma_1), P(\sigma_2), P(\sigma_3)$ as shown in Figure~\ref{fig:lc-variability}, we compute the compliance fraction by integrating over those posteriors using 
\begin{equation}
\label{average compliance}
\mathscr{C}(\text{model}) = \int d\sigma_1d\sigma_2d\sigma_3P(\sigma_1)P(\sigma_2)P(\sigma_3) \mathscr{F}(\sigma_1,\sigma_2,\sigma_3).
\end{equation}
The compliance fraction denotes the expectation value of $\mathscr{F}(\sigma_1,\sigma_2,\sigma_3)$ given the probability distributions of $\sigma_1$,$\sigma_2$, and $\sigma_3$. 

We compute this compliance fraction for black hole masses in the range $5 \times 10^9 M_\odot \leq M_{\rm{BH}} \leq 9.5 \times 10^9 M_\odot$ and use the maximum number as the compliance fraction for a given GRMHD model.

\section{Comparison of GRMHD Simulations to M87 Observations}
\label{sec:obs_comp}

\begin{figure*}[t!]
	\centerline{\includegraphics[width = 0.5\textwidth]{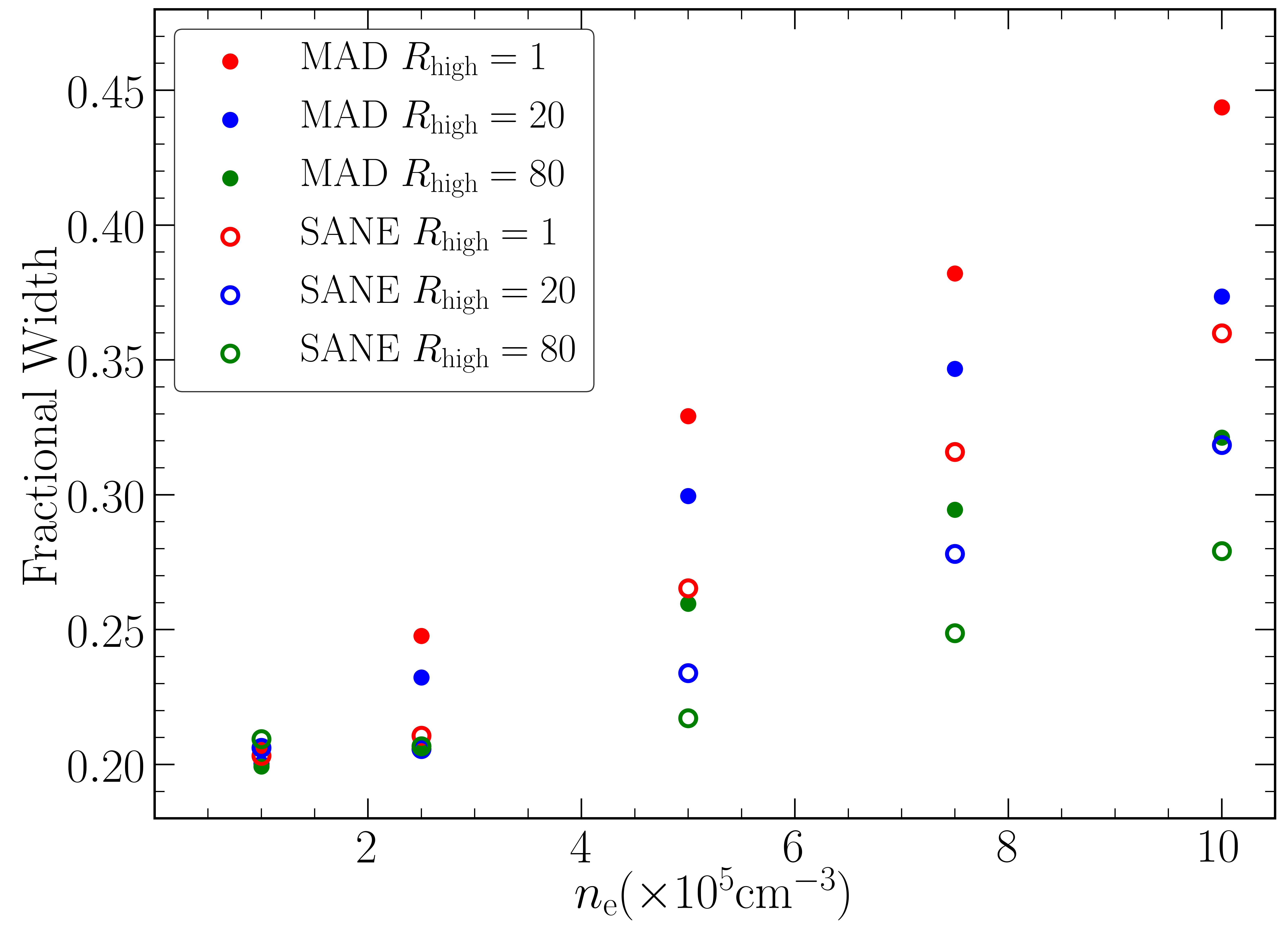}
				\includegraphics[width = 0.5\textwidth]{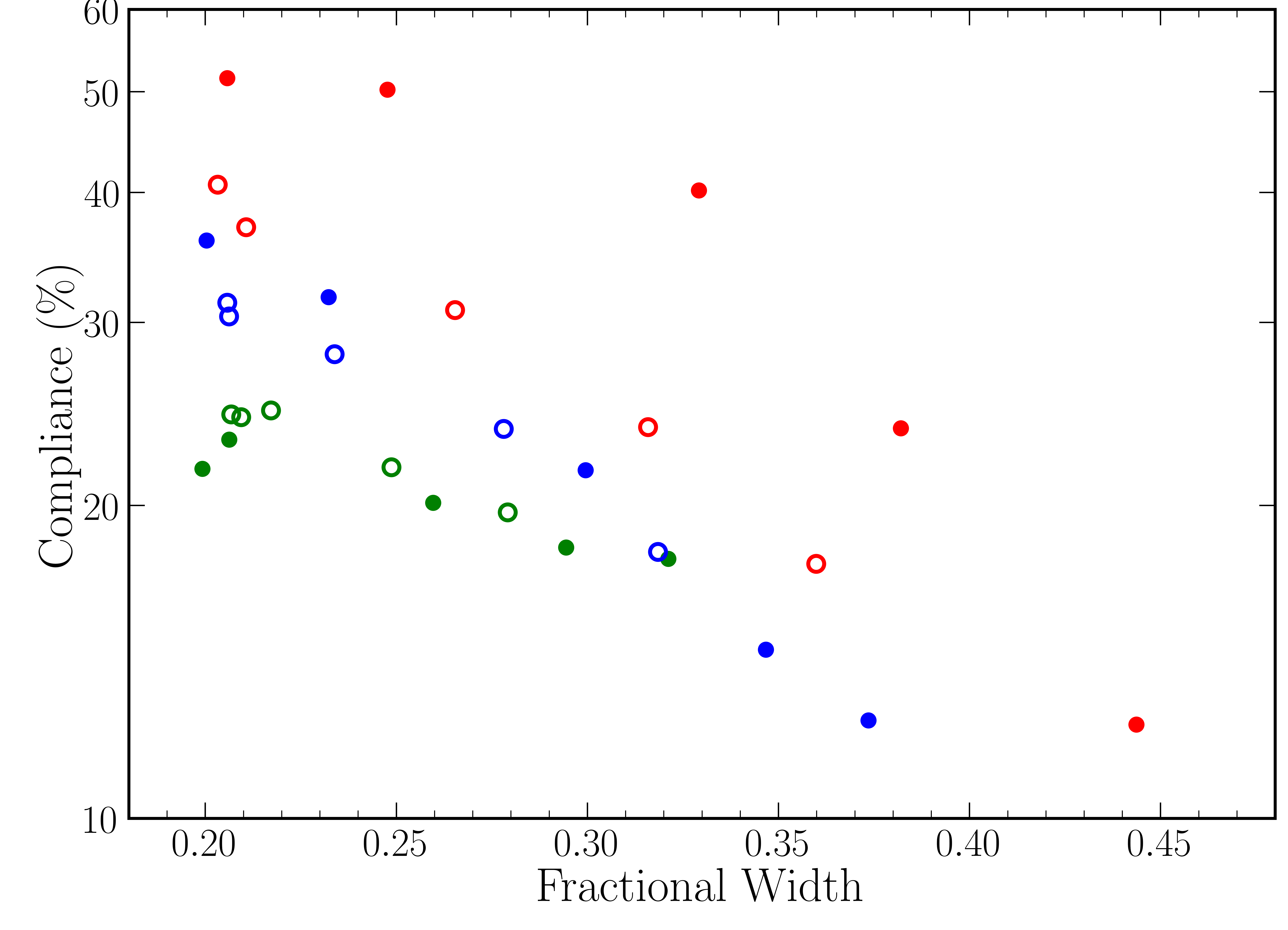} }
\caption{\textit{Left} - The fractional width of the ring of emission in the average image of each simulation in set A. \textit{Right} - The anticorrelation between fractional width and compliance fraction in the simulation set A. }
\label{fig:setA_compl}
\end{figure*}

\begin{table}[t]
\caption{Compliance fraction for Set A}
\label{tab:probabilities_ne}
\begin{threeparttable}
\begin{tabular}{c c c c c}
\toprule
\multirow{2}{*}{Model} &  \multirow{2}{*}{ $n_{\rm{e}}$~($\times10^5 \rm{cm}^{-3}$) }  & \multicolumn{3}{c}{$R_{\rm{high}}$} \\
\cmidrule{3-5}
 &  & 1 & 20 & 80 \\
\midrule
\multirow{5}{*}{SANE\ a = +0.9} & $1$ &  $41\% $&  30\% &  $24 \%$\\
 & $2.5$ &  37\% &  31\% &  24\% \\
 & $5 $&  $31\% $&$  28 \% $&  $25 \% $\\
 & $7.5$ &  $24 \% $& $24 \% $&$  22 \% $\\
 & $10$ &  $18 \% $& $18 \% $& $20 \% $\\
\midrule
\multirow{5}{*}{MAD\ a = +0.9} & $1$ &  52\% &  36\%  &  22\% \\
 & $2.5$ &  50\% & 32\% &  23\% \\
 & $5 $&  40\% &  22\% &  20\% \\
 & $7.5$ & 24\% &  15\% &  18\% \\
 & $10$ & 12\% &  12\% &  18\% \\
 \bottomrule
 \end{tabular}
\end{threeparttable}
\end{table}

\begin{table*}[t]
\caption{Compliance fraction for Set B}
\label{tab:probabilities}
\begin{center}
\begin{threeparttable}

\newcolumntype{C}{>{$}c<{$}}
\begin{tabular}{C C C C C C C C}
\toprule
\multirow{2}{*}{Model} & \multirow{2}{*}{Spin} & & & \multicolumn{2}{C}{R_{\rm{high}}}& & \\
\cmidrule{3-8}
 &  & 1 & 10 & 20 & 40 & 80 & 160 \\
\midrule
\multirow{5}{*}{SANE} & -0.94 & <1\% \tnote{abc} & 2\% \tnote{abc} & 3\%\tnote{abc} & 2\%\tnote{abc} & 1\%\tnote{abc} & <1\%\tnote{abc} \\
{} & -0.5 & <1\%\tnote{a} & 12\%\tnote{a} & 12\%\tnote{ab} & 12\%\tnote{ab} & 8\%\tnote{ab} & 7\%\tnote{ab} \\
{} & 0.0 & <1\%\tnote{ab} & 3\%\tnote{ab} & 8\%\tnote{a} & 11\%\tnote{ab} & 11\%\tnote{ab} & 10\%\tnote{ab} \\
{} & +0.5 & 6\%\tnote{ab} & 2\%\tnote{ab} &  2\%\tnote{ab} & 6\%\tnote{ab} & 10\%\tnote{ab} & 10\%\tnote{ab} \\
{} & +0.94 & 21\%\tnote{b} & 21\%\tnote{b} & 15\%\tnote{ab} & 15\%\tnote{ab} & 19\%\tnote{abc} & 24\%\tnote{abc} \\
\midrule
\multirow{5}{*}{MAD}  & -0.94 & 7\%\tnote{bc} & 6\%\tnote{abc}  & 7\%\tnote{abc}& 6\%\tnote{abc} & 6\%\tnote{abc} & 9\%\tnote{abc} \\
{} & -0.5 & 10\%\tnote{b} &12\%\tnote{ab} & 13\% \tnote{abc}&11\% \tnote{abc}& 11\%\tnote{abc} &11\%\tnote{abc} \\
{} & 0.0 & 12\% \tnote{b}& 11\% \tnote{ab}& 15\%\tnote{ab} & 13\% \tnote{ab}& 10\%\tnote{ab} & 9\%\tnote{ab} \\
{} & +0.5 & 11\%\tnote{b} & 16\% \tnote{abc}& 18\%\tnote{abc} & 19\%\tnote{abc} & 18\%\tnote{abc} & 17\% \tnote{abc}\\
{} & +0.94 & 20\%\tnote{c} & 24\% \tnote{bc}& 24\%\tnote{abc} & 20\%\tnote{abc} & 16\%\tnote{abc} & 12\%\tnote{abc} \\
\bottomrule
\end{tabular}
\begin{tablenotes}
\item \textbf{Notes}
\item[a] Satisfies Radiative Efficiency constraint.
\item[b] Satisfies X-Ray Luminosity constraint.
\item[c] Satisfies Jet Power constraint.
\end{tablenotes}
\end{threeparttable}
\end{center}
\end{table*}

In this section, we use the two sets of simulations we discuss in \S\ref{sec:GRMHD} to explore the dependence of the compliance fraction on the magnetic field configuration, electron temperature prescription in the plasma, and electron-density scale (set~A), as well as on the black-hole spin (set~B). These two sets are meant to serve as two slices in the large parameter space of simulations.

Table~\ref{tab:probabilities_ne} shows the compliance fractions calculated for the 30 simulations in Set~A. Overall, these compliance fractions are rather small, indicating that a lot of simulations show more frequent periods of high variability, even in the triangles that observationally do not exhibit large changes in closure phase across the 6 days of the campaign. The dominant distinguishing characteristic of the simulations that show a higher compliance fraction is the small value for the electron density scale. There is smaller difference in compliance fractions between the MAD and SANE simulations. We observe that for small values of $R_{\rm high}$, the dependence of compliance fraction on the fractional width is stronger.

We explored how the electron density scale affects the image structure and variability. To this end, we calculated the characteristic properties of the average image in each simulation, using the image-domain techniques discussed in \citet{2019ApJ...875L...4E} and \citet{2019ApJ...875L...6E}. We measured the fractional width of each average image $\delta w_f$ as the ratio of the width $\delta w$ to the ring diameter $d$, i.e. $\delta w_f \equiv \delta w / d$. We found the fractional widths of the images correlate strongly with the electron density scale, as shown in Figure~\ref{fig:setA_compl}. This is expected, given that increasing the electron density scale also increases the emissivity and optical depth in the accretion flow, as shown in the top panels of Fig.~\ref{fig:ne_heatmaps}~\citep{2015ApJ...799....1C}.

The right panel of Figure~\ref{fig:setA_compl} shows indeed that the compliance fraction is anticorrelated with the fractional width of the average images of the simulations. When the bright emission ring in an image is narrow, it traces the outline of the black hole shadow and, therefore, its shape and brightness distribution is determined primarily by gravitational lensing effects. On the other hand, when the bright emission ring is broad, variability due to the turbulence in the accretion flow introduces more apparent effects. The low level of variability across the 6 days observed in M87 argue in favor of images characterized by thin ($\sim 20$\%) rings of emission. 

In the simulations of Set~B, the electron number-density scale was tuned to generate a fixed source flux of $0.5$~Jy. Different configurations were explored with both rotating and counterrotating black-hole spins of various magnitudes. The compliance fractions are shown in Table~\ref{tab:probabilities} and are, overall, lower than those of the Set~A simulations. As in the previous case, the MAD simulations have on average a higher compliance fraction than the SANE simulations and there is only a marginal dependence on the electron temperature parameter $R_{\rm high}$. However, there is a significant dependence on the spin of the black hole, with co-rotating, high-spin simulations producing the highest compliance fractions. 

The table also shows constraints imposed on the models of Set~B by other considerations, not originating from the EHT observations~\citep{2019ApJ...875L...5E}. They include constraints related to {\em (a)} the simulations having achieved radiative equilibrium, {\em (b)} the predicted X-ray flux to be $L_X = (4.4 \pm 0.1) \times 10^{40} \rm{erg}\ \rm{s}^{-1}$, as measured during the nearly simultaneous observations with  the \textit{Chandra} X-Ray Observatory and the Nuclear Spectroscopic Telescope Array (\textit{NuSTAR}), and {\em (c)} the jet power being in range $10^{42}$ to $10^{45} \rm{erg}\ \rm{s}^{-1}$. Simulations with fast black-hole spins, which are consistent with these constraints, are also characterized by larger compliance fractions, giving preference to these models.

\section{Discussion}

In this paper, we studied the variability of the black-hole image structure of M87 at timescales comparable to the fastest dynamical timescale near the horizon, as it is imprinted on the closure phases measured during the EHT 2017 observations. We identified three linearly independent closure triangles that exhibit a persistent evolution pattern of closure phases over the course of each night of observing but show little variation around this pattern across the six days of observations, namely, ALMA-LMT-SMT, ALMA-PV-SMT, and ALMA-PV-LMT. In other words, the closure phase evolution in these triangles follows set tracks determined by the rotation of the baselines on the $u-v$ plane but shows very little scatter around these tracks from day to day. We quantified the degree of variability in these triangles to be $\sim3-5^\circ$. This inferred level of variability is comparable to the $\sim 2^\circ$ systematic error in measurement of closure phases \citep{2019ApJ...875L...3E}. 

We also found three triangles that exhibit a high level of day-to-day variability, namely, ALMA-SMA-LMT, ALMA-SMA-SMT, and ALMA-PV-SMA. The change in closure phase across 6 days in a given location on the $u-v$ plane for these triangles can be $\sim 90-180^\circ$. We identified them as triangles with at least one baseline that encounters a deep visibility minimum on the $u-v$ plane. This is in agreement with expectations based on theoretical models that reveal the presence of highly variable but highly localized regions on the $u-v$ plane associated with the locations of these minima. 

We used GRMHD simulations to explore the dependence of closure-phase variability on different model parameters and at different locations in the $u-v$ plane. We found that the most discriminating image characteristic of models in terms of the degree of closure phase variability is the fractional width of the ring of high intensity on the image. Models that best reproduce the observed small level of variability are those with thin ring-like images with structures dominated by gravitational lensing effects and thus least affected by turbulence in the accreting plasmas.

Among the models we explored, there is marginal difference between SANE and MAD simulations, which explore different magnetic field configurations in the accretion flows, with some preference for the MAD models. There is also a small dependence on black-hole spin, with high-spin co-rotating models showing the lowest level of day-to-day variability, in agreement with the observations. 

These findings demonstrate that the method we introduced to quantify day-to-day variability in closure phase data and compare it to models is a useful tool in exploring the origin of variability in horizon-scale images of black holes and in discriminating between models.

\acknowledgements

This work was supported in part by NSF PIRE award 1743747 and NASA ATP award 80NSSC20K0521. L. M. acknowledges support from an NSF Astronomy and Astrophysics Postdoctoral Fellowship under award no.\ AST-1903847. M.W. acknowledges the support of the Black Hole Initiative at Harvard University, which is funded by grants from the John Templeton Foundation and the Gordon and Betty Moore Foundation to Harvard University. All ray tracing calculations for Set~A were performed with the El Gato GPU cluster at the University of Arizona that is funded by NSF award 1228509. All analyses for Set~B were performed on CyVerse, supported by NSF grants DBI-0735191, DBI-1265383, and DBI-1743442. 

The Event Horizon Telescope Collaboration thanks the following
organizations and programs: the Academy
of Finland (projects 274477, 284495, 312496, 315721); the Agencia Nacional de Investigación y Desarrollo (ANID), Chile via NCN$19\_058$ (TITANs) and Fondecyt 3190878, the Alexander
von Humboldt Stiftung; an Alfred P. Sloan Research Fellowship;
Allegro, the European ALMA Regional Centre node in the Netherlands, the NL astronomy
research network NOVA and the astronomy institutes of the University of Amsterdam, Leiden University and Radboud University;
the Black Hole Initiative at
Harvard University, through a grant (60477) from
the John Templeton Foundation; the China Scholarship
Council;  Consejo
Nacional de Ciencia y Tecnolog\'{\i}a (CONACYT,
Mexico, projects  U0004-246083, U0004-259839, F0003-272050, M0037-279006, F0003-281692,
104497, 275201, 263356);
the Delaney Family via the Delaney Family John A.
Wheeler Chair at Perimeter Institute; Dirección General
de Asuntos del Personal Académico-—Universidad
Nacional Autónoma de México (DGAPA-—UNAM,
projects IN112417 and IN112820); the European Research Council Synergy
Grant "BlackHoleCam: Imaging the Event Horizon
of Black Holes" (grant 610058); the Generalitat
Valenciana postdoctoral grant APOSTD/2018/177 and
GenT Program (project CIDEGENT/2018/021); MICINN Research Project PID2019-108995GB-C22;
the
Gordon and Betty Moore Foundation (grant GBMF-3561); the Istituto Nazionale di Fisica
Nucleare (INFN) sezione di Napoli, iniziative specifiche
TEONGRAV; the International Max Planck Research
School for Astronomy and Astrophysics at the
Universities of Bonn and Cologne; 
Joint Princeton/Flatiron and Joint Columbia/Flatiron Postdoctoral Fellowships, research at the Flatiron Institute is supported by the Simons Foundation; 
the Japanese Government (Monbukagakusho:
MEXT) Scholarship; the Japan Society for
the Promotion of Science (JSPS) Grant-in-Aid for JSPS
Research Fellowship (JP17J08829); the Key Research
Program of Frontier Sciences, Chinese Academy of
Sciences (CAS, grants QYZDJ-SSW-SLH057, QYZDJSSW-
SYS008, ZDBS-LY-SLH011); the Leverhulme Trust Early Career Research
Fellowship; the Max-Planck-Gesellschaft (MPG);
the Max Planck Partner Group of the MPG and the
CAS; the MEXT/JSPS KAKENHI (grants 18KK0090,
JP18K13594, JP18K03656, JP18H03721, 18K03709,
18H01245, 25120007); the Malaysian Fundamental Research Grant Scheme (FRGS)\\ FRGS/1/2019/STG02/UM/02/6; the MIT International Science
and Technology Initiatives (MISTI) Funds; the Ministry
of Science and Technology (MOST) of Taiwan (105-
2112-M-001-025-MY3, 106-2112-M-001-011, 106-2119-
M-001-027, 107-2119-M-001-017, 107-2119-M-001-020,
107-2119-M-110-005, 108-2112-M-001-048, and 109-2124-M-001-005); the National Aeronautics and
Space Administration (NASA, Fermi Guest Investigator
grant 80NSSC20K1567, NASA Astrophysics Theory Program grant 80NSSC20K0527, NASA NuSTAR award 80NSSC20K0645); the National
Institute of Natural Sciences (NINS) of Japan; the National
Key Research and Development Program of China
(grant 2016YFA0400704, 2016YFA0400702); the National
Science Foundation (NSF, grants AST-0096454,
AST-0352953, AST-0521233, AST-0705062, AST-0905844, AST-0922984, AST-1126433, AST-1140030,
DGE-1144085, AST-1207704, AST-1207730, AST-1207752, MRI-1228509, OPP-1248097, AST-1310896,  AST-1555365, AST-1615796, AST-1715061, AST-1716327,  AST-1903847,AST-2034306); the Natural Science
Foundation of China (grants 11573051, 11633006,
11650110427, 10625314, 11721303, 11725312, 11933007, 11991052, 11991053); a fellowship of China Postdoctoral Science Foundation (2020M671266); the Natural
Sciences and Engineering Research Council of
Canada (NSERC, including a Discovery Grant and
the NSERC Alexander Graham Bell Canada Graduate
Scholarships-Doctoral Program); the National Youth
Thousand Talents Program of China; the National Research
Foundation of Korea (the Global PhD Fellowship
Grant: grants NRF-2015H1A2A1033752, 2015-
R1D1A1A01056807, the Korea Research Fellowship Program:
NRF-2015H1D3A1066561, Basic Research Support Grant 2019R1F1A1059721); the Netherlands Organization
for Scientific Research (NWO) VICI award
(grant 639.043.513) and Spinoza Prize SPI 78-409; the
New Scientific Frontiers with Precision Radio Interferometry
Fellowship awarded by the South African Radio
Astronomy Observatory (SARAO), which is a facility
of the National Research Foundation (NRF), an
agency of the Department of Science and Technology
(DST) of South Africa; the Onsala Space Observatory
(OSO) national infrastructure, for the provisioning
of its facilities/observational support (OSO receives
funding through the Swedish Research Council under
grant 2017-00648) the Perimeter Institute for Theoretical
Physics (research at Perimeter Institute is supported
by the Government of Canada through the Department
of Innovation, Science and Economic Development
and by the Province of Ontario through the
Ministry of Research, Innovation and Science);  the Spanish
Ministerio de Economía y Competitividad (grants
PGC2018-098915-B-C21, AYA2016-80889-P, PID2019-108995GB-C21); the State
Agency for Research of the Spanish MCIU through
the "Center of Excellence Severo Ochoa" award for
the Instituto de Astrofísica de Andalucía (SEV-2017-
0709); the Toray Science Foundation; the Consejería de Economía, Conocimiento, Empresas y Universidad of the Junta de Andalucía (grant P18-FR-1769), the Consejo Superior de Investigaciones Científicas (grant 2019AEP112);
the US Department
of Energy (USDOE) through the Los Alamos National
Laboratory (operated by Triad National Security,
LLC, for the National Nuclear Security Administration
of the USDOE (Contract 89233218CNA000001);
 the European Union’s Horizon 2020
research and innovation programme under grant agreement
No 730562 RadioNet; ALMA North America Development
Fund; the Academia Sinica; Chandra DD7-18089X and TM6-
17006X; the GenT Program (Generalitat Valenciana)
Project CIDEGENT/2018/021. This work used the
Extreme Science and Engineering Discovery Environment
(XSEDE), supported by NSF grant ACI-1548562,
and CyVerse, supported by NSF grants DBI-0735191,
DBI-1265383, and DBI-1743442. XSEDE Stampede2 resource
at TACC was allocated through TG-AST170024
and TG-AST080026N. XSEDE JetStream resource at
PTI and TACC was allocated through AST170028.
The simulations were performed in part on the SuperMUC
cluster at the LRZ in Garching, on the
LOEWE cluster in CSC in Frankfurt, and on the
HazelHen cluster at the HLRS in Stuttgart. This
research was enabled in part by support provided
by Compute Ontario (http://computeontario.ca), Calcul
Quebec (http://www.calculquebec.ca) and Compute
Canada (http://www.computecanada.ca). We thank
the staff at the participating observatories, correlation
centers, and institutions for their enthusiastic support.
This paper makes use of the following ALMA data:
ADS/JAO.ALMA\#2016.1.01154.V. ALMA is a partnership
of the European Southern Observatory (ESO;
Europe, representing its member states), NSF, and
National Institutes of Natural Sciences of Japan, together
with National Research Council (Canada), Ministry
of Science and Technology (MOST; Taiwan),
Academia Sinica Institute of Astronomy and Astrophysics
(ASIAA; Taiwan), and Korea Astronomy and
Space Science Institute (KASI; Republic of Korea), in
cooperation with the Republic of Chile. The Joint
ALMA Observatory is operated by ESO, Associated
Universities, Inc. (AUI)/NRAO, and the National Astronomical
Observatory of Japan (NAOJ). The NRAO
is a facility of the NSF operated under cooperative agreement
by AUI. APEX is a collaboration between the
Max-Planck-Institut f{\"u}r Radioastronomie (Germany),
ESO, and the Onsala Space Observatory (Sweden). The
SMA is a joint project between the SAO and ASIAA
and is funded by the Smithsonian Institution and the
Academia Sinica. The JCMT is operated by the East
Asian Observatory on behalf of the NAOJ, ASIAA, and
KASI, as well as the Ministry of Finance of China, Chinese
Academy of Sciences, and the National Key R\&D
Program (No. 2017YFA0402700) of China. Additional
funding support for the JCMT is provided by the Science
and Technologies Facility Council (UK) and participating
universities in the UK and Canada. The
LMT is a project operated by the Instituto Nacional
de Astrófisica, Óptica, y Electrónica (Mexico) and the
University of Massachusetts at Amherst (USA). The
IRAM 30-m telescope on Pico Veleta, Spain is operated
by IRAM and supported by CNRS (Centre National de
la Recherche Scientifique, France), MPG (Max-Planck-
Gesellschaft, Germany) and IGN (Instituto Geográfico
Nacional, Spain). The SMT is operated by the Arizona
Radio Observatory, a part of the Steward Observatory
of the University of Arizona, with financial support of
operations from the State of Arizona and financial support
for instrumentation development from the NSF.
The SPT is supported by the National Science Foundation
through grant PLR- 1248097. Partial support is
also provided by the NSF Physics Frontier Center grant
PHY-1125897 to the Kavli Institute of Cosmological
Physics at the University of Chicago, the Kavli Foundation
and the Gordon and Betty Moore Foundation grant
GBMF 947. The SPT hydrogen maser was provided on
loan from the GLT, courtesy of ASIAA. The EHTC has
received generous donations of FPGA chips from Xilinx
Inc., under the Xilinx University Program. The EHTC
has benefited from technology shared under open-source
license by the Collaboration for Astronomy Signal Processing
and Electronics Research (CASPER). The EHT
project is grateful to T4Science and Microsemi for their
assistance with Hydrogen Masers. This research has
made use of NASA’s Astrophysics Data System. We
gratefully acknowledge the support provided by the extended
staff of the ALMA, both from the inception of
the ALMA Phasing Project through the observational
campaigns of 2017 and 2018. We would like to thank
A. Deller and W. Brisken for EHT-specific support with
the use of DiFX. We acknowledge the significance that
Maunakea, where the SMA and JCMT EHT stations
are located, has for the indigenous Hawaiian people.

\appendix

\section{Approximate Degeneracy of the Radiative Transfer Solution}
\label{sec:appendix 1}

In this Appendix, we present analytical arguments to show that the mm-wavelength images calculated from GRMHD simulations of accretion flows with parameters relevant to the M87 black hole are approximately invariant to the product $n_e^2 M_{\rm{BH}}$, where $n_e$ is the electron density scale and $M_{\rm{BH}}$ is the mass of the black hole.  

The radiative transfer equation at a frequency $\nu$ is given by  
\begin{align}
\label{eq:radiative transfer 1}
\frac{dI_\nu}{ds} &= \eta_\nu - \chi_\nu I_\nu\;,
\end{align}
where $I_\nu$ denotes the specific intensity at a frequency $\nu$, $s$ denotes the distance travelled along the path of a photon, $\eta_\nu$ denotes the emissivity of the medium, and $\chi_\nu$ denotes the opacity of the medium. Equation~(\ref{eq:radiative transfer 1}) can be recast as 
\begin{equation}
\label{eq:radiative transfer 2}
\frac{dI_\nu}{ds} =\chi_\nu\left(\frac{\eta_\nu}{\chi_\nu} - I_\nu \right). 
\end{equation}
The fraction $\eta_\nu/\chi_\nu$ is the source function $S_\nu$ and, for thermal emission, is equal to the blackbody function, which depends solely on temperature $T$. We denote this as $\eta_\nu/\chi_\nu \equiv \mathscr{B}_\nu(T)$. \\

Throughout this work, we use the analytic approximation devised by~\citet{Leung_2011} for synchrotron emissivity from a thermal distribution of relativistic electrons 
\begin{equation}
    \eta_\nu=n_e \frac{\sqrt{2}\pi e^2 \nu_{\rm s}}{3 K_2(1/\Theta_e)c}
    \left(X^{1/2}+2^{11/12}X^{1/6}\right)^2 \exp(-X^{1/3})\;,
\end{equation}
where $X\equiv \nu/\nu_{\rm s}$, $\nu_{\rm s}\equiv (2/9)\nu_{\rm c} \Theta_e^2\sin\theta$, $\nu_c\equiv e B/(2\pi m_e c)$, $\Theta_e\equiv k_{\rm b} T_{\rm e}/(m_{\rm e} c^2)$ is the dimensionless electron temperature, and $K_2$ denotes the modified Bessel function of the second kind. In the these expressions, $\theta$ denotes the angle between the magnetic field and the emitted photon, $B$ denotes the magnetic field strength, and $T_e$ denotes the electron temperature. We use the constants $m_e$, $e$, $k_b$, and $c$ to denote the electron mass, electron charge, the Boltzmann's constant, and the speed of light respectively. It is explicit in this expression that the synchrotron emissivity and opacity, for a given electron temperature and magnetic field, scale proportionally to the electron number density $n_{\rm{e}}$. We show here that, for the parameters of the black hole at the center of the M87 galaxy, the synchrotron opacity and emissivity at a wavelength of $\lambda=1.3$~mm also scale proportionally to a power of the magnetic field, i.e., $\propto B^\alpha$, with $\alpha\sim 2$.

We estimate the properties of the plasma in the inner accretion flow of M87 using the following analytic model. The electron density at a given radius $r$ is given by the continuity equation, assuming a spherical accretion rate $\dot{M}$ via the relation
\begin{equation}
    \dot{M}=4\pi r^2 \left(\frac{h}{r}\right) m_p n_e u^r\;,
\end{equation}
where $h/r$ is the scale height of the accretion flow, $m_p$ is the proton mass (assuming fully ionized hydrogen), and $u^r$ is the radial component of the accretion flow. We take the latter to be a fraction $\xi$ of the free-fall velocity, i.e.,
\begin{equation}
    u^r=\xi \left(\frac{GM_{\rm{BH}}}{r}\right)^{1/2}\;.
\end{equation}
We also express the accretion rate as $\dot{M}\equiv\dot{m}\dot{M}_{\rm E}$, in terms of the Eddington accretion rate defined by
\begin{equation}
    \dot{M}_{\rm E}\equiv \frac{L_{\rm E}}{\epsilon c^2}=\frac{4 \pi G M_{\rm{BH}} m_{\rm p}}{\epsilon c \sigma_{\rm T}}\;,
\end{equation}
where $L_{\rm E}$ is the Eddington critical luminosity, $\epsilon$ is the radiative efficiency of the flow, $\sigma_{\rm T}$ is the Thomson cross section, and $\dot{m}$ is the mass accretion rate in units of the Eddington accretion rate. Under these conditions, the electron density is
\begin{eqnarray}
n_e&=&\left(\frac{c^2}{GM_{\rm{BH}} \sigma_{\rm T}}\right)\dot{m}\epsilon^{-1}\xi^{-1}\left(\frac{h}{r}\right)^{-1}
\left(\frac{r c^2}{GM_{\rm{BH}}}\right)^{-3/2}\nonumber\\
&=&7\times 10^5 \left(\frac{M_{\rm{BH}}}{6.5\times 10^9 M_\odot}\right)^{-1} \left(\frac{\dot{m}}{2\times 10^{-5}}\right)
\left(\frac{r c^2}{5 GM_{\rm{BH}}}\right)^{-3/2}~{\rm cm}^{-3}\;.
\end{eqnarray}
where in the last expression we have used typical values for the M87 black hole and set $\xi=\epsilon=0.1$ and $h/r=0.4$.

Because of the radiatively inefficient character of the accretion flow around the M87 black hole, the ion temperature $T_i$ is a fraction of the virial temperature
\begin{equation}
    T_{\rm v}=\frac{GM_{\rm{BH}} m_p}{3 k_{\rm B} r}\;.
\end{equation}
It is also customary to write the electron temperature as a fraction $1/R$ of the ion temperature, i.e.,
\begin{equation}
    T_{\rm e}=\frac{1}{R} \left(\frac{G M_{\rm{BH}} m_{\rm p}}{3 k_B r}\right)
    \left(\frac{T_i}{T_{\rm v}}\right)\;.
\end{equation}
Finally, we write the magnetic field in terms of the plasma $\beta$ parameter as
\begin{equation}
n_e k_{\rm B}(T_i+T_e)=\beta\frac{B^2}{8\pi}
\end{equation}
such that
\begin{eqnarray}
B&=&2c^2\left(\frac{2\pi m_{\rm p}}{3 G M_{\rm{BH}} \sigma_{\rm T}}\right)^{1/2}
\left[\frac{\dot{m}(1+R)}{\epsilon \xi \beta R}\right]^{1/2}
\left(\frac{T_i}{T_{\rm v}}\right)^{1/2}
\left(\frac{h}{r}\right)^{-1/2}
\left(\frac{rc^2}{GM_{\rm{BH}}}\right)^{-5/4}\nonumber\\
&=&9 \left(\frac{M_{\rm{BH}}}{6.5\times 10^9 M_\odot}\right)^{-1/2} \left(\frac{\dot{m}}{2\times 10^{-5}}\right)^{1/2}
\left(\frac{r c^2}{5 GM_{\rm{BH}}}\right)^{-5/4}~{\rm G}\;,
\end{eqnarray}
and we have set $T_i/T_{\rm v}=1/3$ and $\beta=10$\;.

\begin{figure}[t]
	\centerline{\includegraphics[width = 0.5\textwidth]{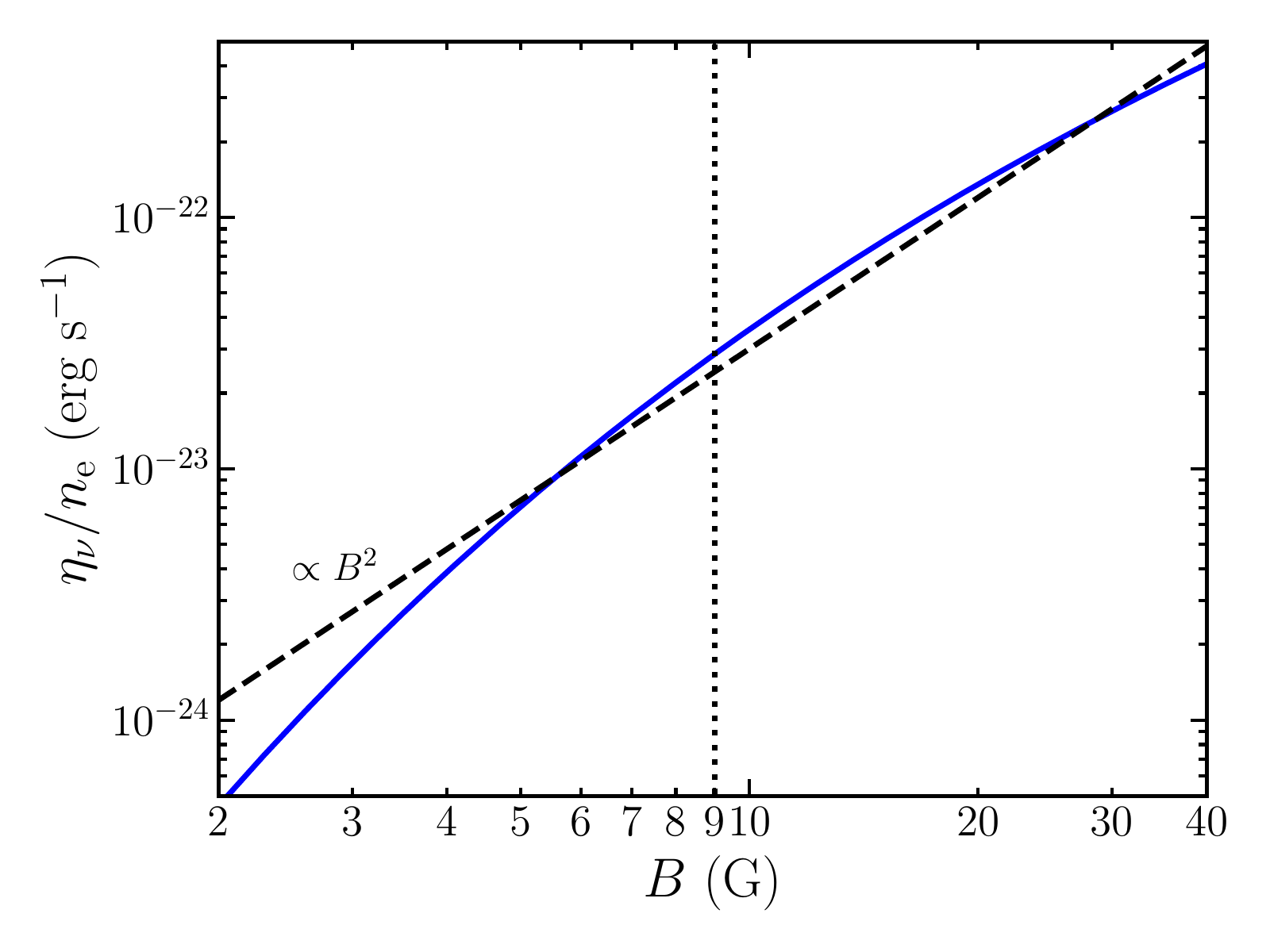}}
\caption{The ratio of the thermal synchrotron emissivity $\eta_\nu$ to the electron density $n_e$, as a function of the strength of the magnetic field, evaluated at an observing frequency of the EHT, $\nu=230$~GHz, and for parameters appropriate to the inner accretion flow in the black hole at the center of the M87 galaxy (see text). The vertical dashed line shows the inferred strength of the magnetic field near the black-hole horizon. The dashed line shows the approximate $\sim B^2$ scaling. }
\label{fig:synch}
\end{figure}

Figure~\ref{fig:synch} shows the ratio $\eta_\nu/n_e$ evaluated at the observed frequency of the EHT ($\nu=230$~GHz, $\lambda=1.33$~mm) and at a radial distance $r=5GM_{\rm{BH}}/c^2$, as a function of the strength of the magnetic field $B$, for the values of the various parameters used in the previous equations, i.e., for the conditions of the inner accretion flow around the black hole at the center of M87. This figure demonstrates that, over a broad range of magnetic field strengths (spanning more than an order of magnitude), the synchrotron emissivity scales approximately as $\eta_\nu\sim n_{\rm e}B^\alpha$, with $\alpha\simeq 2$. This allows us to write the emissivity as
\begin{equation}
\eta_\nu \simeq n_{\rm e}B^\alpha {\cal B}_\nu(T) f_\nu(T,\vec{r}), 
\label{eq:emiss}
\end{equation}
and the opacity as
\begin{equation}
    \chi_\nu\simeq n_{\rm{e}}B^\alpha f_\nu(T,\vec{r})\;,
    \label{eq:opac}
\end{equation}
where $f_\nu(T,\vec{r})$ captures the scaling of the opacity with temperature $T$ and location in the accretion flow (see similar arguments in \citealt{2015ApJ...799....1C}). 

In order to calculate the various simulated images, we used the plasma properties from long GRMHD simulations. Nonradiative GRMHD simulations are invariant to a rescaling of the density by a factor ${\cal M}$, as long as the magnetic field also is also rescaled by a factor of ${\cal M}^{1/2}$ and the internal energy by a factor of ${\cal M}$. In other words, $B\sim n_{\rm e}^{1/2}$ and the Alfv\'en speed $B/(m_{\rm p}n_e)^{1/2}$ is the quantity that remains invariant under rescaling. Combined with the approximate expressions~(\ref{eq:emiss})-(\ref{eq:opac}) derived above, this implies that the synchrotron emissivity and opacity evaluated using the simulations outputs are invariant to rescaling, as long as the product $n_e B^\alpha\sim n_e^{1+\alpha/2}$ remains constant. 

Finally, the integration of the transfer equation is performed on a coordinate system in which the distances are expressed in terms of the length scale set by the mass of the black hole, i.e.,
\begin{equation}
\label{eq:dimless length}
ds \equiv ds' \left(\frac{GM_{\rm{BH}}}{c^2}\right)\;.
\end{equation}
Rewriting equation~(\ref{eq:radiative transfer 2}) using the scaling~(\ref{eq:emiss})-(\ref{eq:dimless length}), we find 
\begin{align}
\label{eq:radiative transfer 3}
\frac{dI}{ds'} &= n_{\rm{e}}^{1+\alpha/2} f(T,\vec{r}) \frac{GM_{\rm{BH}}}{c^2}\left[\mathscr{B}(T) - I\right] \\
&= \left(n_{\rm{e}}^{1+\alpha/2} M_{\rm{BH}}\right)\frac{G}{c^2} f(T,\vec{r}) \left[\mathscr{B}(T) - I\right]. 
\end{align}
We drop the $\nu$ in subscript to indicate quantities calculated at $\nu = 230$ GHz. From this last expression it is apparent that the electron number density $n_{\rm{e}}$ and the black hole mass $M_{\rm{BH}}$ are degenerate quantities in the solution of the radiative transfer problem at wavelengths where the dominant source of opacity is due to synchrotron processes, with a degeneracy in the product $n_{\rm{e}}^{1+\alpha/2} M_{\rm{BH}}$ and with $\alpha\simeq 2$. 

\bibliographystyle{apj}

\bibliography{cp_citations}

\end{document}